\documentclass{aa}  

\usepackage{subcaption}
\usepackage{makecell}
\usepackage{float}
\usepackage{balance}
\usepackage{graphicx}
\usepackage{caption}
\usepackage[export]{adjustbox}
\DeclareCaptionLabelFormat{boldlabel}{\textbf{#1~#2}}
\usepackage{txfonts}
\usepackage[%
    colorlinks=true,
    citecolor=blue,
    pdfborder={0 0 0},
    linkcolor=blue
]{hyperref}

\titlerunning{Investigating AGN feedback in H$\alpha$-luminous galaxy clusters}
\authorrunning{I. Fornasiero et al.}

\begin{document} 

   \title{Investigating AGN feedback in H$\alpha$-luminous galaxy clusters: \\
   first {\it Chandra} X-ray analysis of Abell 2009}


   \author{I. Fornasiero\inst{1,} \inst{2},
          F. Ubertosi\inst{3,} \inst{4}
          \and
          M. Gitti\inst{3,} \inst{4}
          }

   \institute{Scuola Universitaria Superiore IUSS Pavia, Palazzo del Broletto, piazza della Vittoria 15, I-27100 Pavia, Italy\\
        \email{ilaria.fornasiero@iusspavia.it}
         \and
         Department of Physics, University of Trento, Via Sommarive 14, I-38123 Povo (TN), Italy
         \and
             Dipartimento di Fisica e Astronomia, Università di Bologna, Via Gobetti 93/2, 40129 Bologna, Italy
         \and
             Istituto Nazionale di Astrofisica (INAF) – Istituto di Radioastronomia (IRA), via Gobetti 101, I-40129 Bologna, Italy
             }

   \date{Received December 5, 2024; accepted February 22, 2025}

 
  \abstract
  {}
{We analyze the X-ray and radio properties of the galaxy cluster Abell 2009 (z $\sim$ 0.152) to complete the in-depth individual study of a subsample of objects from the ROSAT Brightest Cluster Sample (BCS) with a relatively high X-ray flux and H$\alpha$ line luminosity,  which is a promising diagnostic of the presence of cool gas in the cluster cores. Our aim is to investigate the feedback from the active galactic nucleus (AGN) in the central galaxy and the intracluster medium (ICM) of relaxed clusters.}
   {In this work, we analyze archival data from JVLA and \textit{Chandra} observations. We performed a morphological analysis of both the X-ray emission from the ICM of Abell~2009 and of the radio emission from the AGN in the central galaxy. We also performed a spectral analysis of the X-ray emission, to derive the global properties and radial profiles of the thermal gas.}
   {Our X-ray analysis confirms the expectations, based on the selection criteria, that Abell 2009 is a cool-core system. We estimate a cooling radius of $\sim$88 kpc within which the ICM is radiating away its energy at rates of $L_{\rm cool}\sim4.4\times10^{44} \text{ erg s}^{-1}$. Radio observations of the central galaxy reveal a bright core surrounded by radio lobes on 30 kpc scales, with a symmetric butterfly-shaped morphology. We also present the detection of an extended radio galaxy to the northwest of the central one that is also a cluster member of Abell 2009. Although we did not detect any clear X-ray cavity at the position of the central radio lobes by assuming that their size is comparable, we combined the volume of the lobes with the pressure of the surrounding ICM to derive the work done by the AGN on the gas to inflate them. By estimating a cavity age of about 20 Myr, this corresponds to a mechanical power of $\approx10^{45}\text{ erg s}^{-1}$, which is sufficient to counterbalance the radiative cooling losses in Abell 2009. We finally discuss possible correlations between the global properties of the 18 objects from the BCS selection, finding in particular that the number of outbursts required to counterbalance the radiative ICM losses is linearly anticorrelated with the energetics and power of the outburst.}
   {}

   \keywords{clusters --
                AGN feedback --
                intracluster medium --
                X-ray: radio
               }

   \maketitle
%

\section{Introduction}\label{intro}
Within galaxy clusters, the majority of the baryonic component is in the form of a hot, diffuse gas permeating the cluster's volume, known as the intracluster medium (ICM).
This gas emits strongly in X-rays at temperatures of $T\sim 10^7-10^8 K$ (corresponding to a few keV\footnote{$1 \, \text{keV} \approx 1.16 \times 10^7 K$}), primarily through thermal bremsstrahlung radiation and with typical luminosities around $L_x \sim 10^{44}  
\text{ erg} \, \text{s}^{-1}$ (see \citealt{1986RvMP...58....1S}).\\
Nowadays, galaxy clusters are broadly classified into two main categories: cool-core and non cool-core. Cool-core clusters are characterized by a decreasing temperature profile toward the center and a dense (electron density $n_{\rm e} \sim 10^{-2} \text{ cm}^{-3}$) core. 
On the other hand, non cool-core clusters do not show signs of a lower central temperature and higher central density, and they generally show evidence of recent dynamical activity, such as minor or major mergers (e.g., \citealt{2007PhR...443....1M}, \citealt{2010A&A...510A..83R}).\\
Focusing on cool-core systems, the relevance of radiative losses coming from the emission of the hot ICM can be established by considering its characteristic cooling time.
In the central regions of cool-core clusters, where the cooling time ($t_{\rm cool}\propto kT/n_e$) is typically below a few gigayears, the ICM experiences a sufficiently high cooling rate, leading to the dissipation of particles' energy through X-ray radiation.
Early studies of the cores of galaxy clusters predicted the emergence of a cooling flow of material toward the center of the cluster (see \citealt{1994ARA&A..32..277F}).
Nonetheless, later studies have shown that the net rate of gas actually cooling to low temperatures is only 1 - 10\% of the classical expected cooling flow rate (e.g., \citealt{2001ApJ...557..546D}, \citealt{Peterson_2006}).
Historically, there have been two primary strategies to address the inconsistency of the so-called cooling flow problem. As the gas emits radiation without signs of cooling, either the typical signatures of radiative cooling below 1-2 keV are somehow inhibited (e.g., \citealt{Fabian_2001}, \citealt{Glenn_Morris_2003}, \citealt{Mathews_2003}, \citealt{2022MNRAS.515.3336F}) or a mechanism injecting energy into the ICM must exist to counterbalance the cooling process (e.g., \citealt{Markevitch_2001}, \citealt{Voigt_2002}, \citealt{Domainko_2004}, \citealt{McNamara_2007}, \citealt{2012MNRAS.424..190G}). 
Regarding the latter, it has been recognized that central galaxies -- also known as brightest cluster galaxies (BCGs) -- in cool-core clusters show a prevalence of radio activity \citep{2007MNRAS.379..894B}. 
In $\sim$70\% of cases, the BCGs hosting an active galactic nucleus (AGN) were associated with Fanaroff-Riley type I (FRI) radio galaxies (see \citealt{1974MNRAS.167P..31F} and \citealt{1990AJ.....99...14B}). 
FRI radio galaxies are center-brightened showing more luminous and irregular jets in the central parts and faint lobes at the edges \citep{Banfield_2015}.
From \textit{Chandra} telescope images, the central hot gas in numerous cool-core systems exhibits a nonuniform distribution that closely aligns with the extended lobes of radio emission (e.g., \citealt{2001ApJ...558L..15B}, \citealt{2003MNRAS.344L..43F}). These observations also unveiled different structures within the cores of many clusters, such as shocks, ripples, and abrupt density variations. A comparison with radio images suggested that many of these disturbances are associated with the activity of AGN jets. The propagation of the jets displaces the X-ray-emitting gas in the cluster atmosphere, excavating depressions in the ICM that manifest as cavities in X-ray images (e.g., \citealt{2004ApJ...607..800B}, \citealt{McNamara_2007}, \citealt{Gitti_2012}). These structures are currently regarded as an extremely powerful tool to demonstrate that the mechanical power of the AGN is capable of heating the ICM (e.g., \citealt{2006ApJ...643..120B}, \citealt{2008ApJ...686..859B}, \citealt{2012MNRAS.424..190G}).
Multiple comparisons over the years of the cavity power, P$_{\rm cav}$, and the bolometric X-ray luminosity of the cooling region, L$_{\rm cool}$, have indeed shown a consistent scaling between the two quantities (e.g., \citealt{2004ApJ...607..800B}, \citealt{2006ApJ...652..216R}, \citealt{2012MNRAS.427.3468B}, \citealt{2022MNRAS.516L.101O}, \citealt{2023ApJ...944..216U}).
This suggests that, on average, the mechanical power of the jets is able to offset the radiative cooling losses of the ICM.
As multiple AGN outbursts can occur within a cooling time of the ICM (few hundreds of million years), the exact balance between the AGN heating power and the ICM cooling losses also depends on the duty cycle of the AGN, which is the fraction of time during which the central supermassive black hole (SMBH) is actively driving relativistic jets through its surrounding environment (see \citealt{2008ApJ...687..899R}, \citealt{2012MNRAS.421.1569B}, \citealt{Gitti_2012}).\\
In addition to the effects of AGN activity, sloshing motions are frequently observed in the cores of cool-core clusters. Sloshing arises from gravitational perturbations in the potential wells, typically caused by minor mergers or interactions, and manifests as spiral structures in the X-ray surface brightness and temperature maps (e.g., \citealt{2007PhR...443....1M}, \citealt{2010ApJ...717..908Z}). These motions redistribute the energy within the ICM, which impacts the thermodynamic state of the gas. Sloshing can influence cooling by generating thermal instabilities, which play a key role in regulating the distribution and cooling rates of the ICM. At the same time, it can affect the co-spatiality between the AGN and its fuel reservoirs, potentially destabilizing the AGN feedback cycle. Early studies suggest that sloshing does not break the feedback loop, but may influence the timing and efficiency of the AGN activity (e.g., \citealt{2021ApJ...911...66P}, \citealt{2024ApJ...963....8R}).\\
Several optical (e.g., \citealt{2010ApJ...721.1262M}) and submillimeter (e.g., \citealt{2001MNRAS.328..762E}) observations of BCGs have revealed the presence of warm and cold gas filaments, likely cooling out of the ICM and spatially connected to the core of the BCG, which are thought to ultimately provide the fuel for the SMBH (see \citealt{2016MNRAS.460.1758H}, \citealt{2019A&A...631A..22O}, \citealt{2020NatAs...4...10G}, \citealt{2022PhR...973....1D}). In this context, studying systems with strong H$\alpha$ emission and high X-ray flux is essential to understand the feeding and feedback cycle in cool-core clusters, as these features are typical of systems where both cooling and heating processes are most likely at play (e.g., \citealt{2013A&A...555A..93E}).
The AGN-ICM interaction is, in fact, two-fold: not only do the relativistic jets heat the gas, but the gas itself can provide fuel to sustain the AGN activity.
This two-fold interplay is known as the self-regulated feedback loop (see \citealt{McNamara_2007}, \citealt{2012ARA&A..50..455F}), where hot gas cooling initially triggers the SMBH activity (e.g., \citealt{2023A&A...673A..52U}), leading to relativistic jets depositing mechanical energy into the gas and reducing further cooling; as the jet activity weakens, more efficient cooling is restored (see \citealt{2011MNRAS.411..349G} and \citealt{2015ApJ...811...73L} for numerical simulations).\\

This work aims to complete the study of individual targets from a selection of galaxy clusters that are expected to be exemplary cases of the self-regulated feedback loop. As previously discussed in \cite{2013A&A...555A..93E}, \cite{2019ApJ...885..111P}, \cite{2021ApJ...911...66P}, \cite{2023AandA...670A..23U}, \cite{2024ApJ...963....8R}, the conditions that regulate the interaction between AGN feedback and ICM cooling can be best explored in systems with a relatively high X-ray flux and a BCG that is relatively bright in the H$\alpha$ line (suggesting the
presence of cool gas). The above previous works selected systems from the ROSAT Brightest Cluster Sample (BCS) of 201 galaxy clusters with X-ray fluxes greater than $4.4 \times 10^{-12}$ erg cm$^{-2}$ s$^{-1}$ \citep{1998MNRAS.301..881E} and $H\alpha$ luminosities greater than $10^{40}$ erg s$^{-1}$ \citep{1999MNRAS.306..857C}. The resulting 18 objects are reported in Tab.\ref{samplevalues}, and we refer the reader to \cite{2023AandA...670A..23U} for a short summary on the properties of these systems. We note that the BCS catalog and the work of \citet{1999MNRAS.306..857C} are restricted to the northern sky. This makes the selected systems visible from the Jansky Very Large Array (JVLA), which has been routinely used to investigate the properties of radio galaxies at the center of clusters (e.g., \citealt{2008ApJ...686..859B}). A similar selection in the southern sky, where other radio telescopes such as MeerKAT and Australian Square Kilometer Array Pathfinder (ASKAP) could be used to perform a similar analysis, will be the aim of future works.\\
The majority of the galaxy clusters from this selection (listed in Tab.\ref{samplevalues}) have at least one dedicated study in the literature (e.g, \citealt{2003ApJ...587..619S}, \citealt{2006ApJ...648..164M}, 
\citealt{2011ApJ...737...99B}, \citealt{Pandge_2013}, \citealt{Sonkamble_2015}, \citealt{2016ApJ...832..148V}, \citealt{Russell_2017}), which confirmed the expectation that these objects are representative examples of the interplay between the AGN and the ICM. These dedicated studies typically employed {\it Chandra} observations of the ICM, to spatially resolve the evidences of feedback (e.g., the cavities), and JVLA observations of the central radio galaxy. For example, ZwCl1742 shows evidence of cold fronts and a weak shock in {\it Chandra} data, suggesting the presence of sloshing gas in the cluster core or a metal-rich outflow from the central AGN \citep{2013A&A...555A..93E}. Abell 2495 shows an offset between the BCG and the X-ray emission peak, and a potential sloshing-induced activation of the central AGN, which emphasizes the key role of the sloshing effect on the self-regulated feedback loop (\citealt{2019ApJ...885..111P} and \citealt{2024ApJ...963....8R}). Recent works also include the study of Abell 1668 \citep{2021ApJ...911...66P} and ZwCl 235 \citep{2023AandA...670A..23U}, which show important signs of the AGN feedback cycle regulating the cooling process and the presence of a possible sloshing phenomenon in these galaxy clusters.\\
The only cluster in this selection (see Tab.\ref{samplevalues}) that still lacks a dedicated study is Abell 2009 ($z=0.152$), located at RA: 15:00:19.52 and DEC: +21:22:09.90. 
Its $H\alpha$ luminosity is relatively small ($\approx 6 \times 10^{40}$ erg s$^{-1}$) compared to other well-studied clusters at the top of Tab.\ref{samplevalues}, indicating a small star formation rate (which is also evident from the low upper limit of infrared luminosity given by \citealt{2008ApJS..176...39Q}).\\
\renewcommand{\arraystretch}{1.3}
\begin{table*}[h]
\centering
\caption{Selection of high X-ray flux and high H$\alpha$ luminosity galaxy clusters from the BCS sample.}
\begin{tabular}{ccccccccc}
\hline
  Cluster& \makecell{F$_X$ \\ {[erg cm$^{-2}$ s$^{-1}$]}} & \makecell{L$_{H\alpha}$ \\ {[erg s$^{-1}$]}} & 
\makecell{$E_{\text{cav}}$ \\ {[10$^{60}$ erg]}} & 
\makecell{$t_{\text{cav}}$ \\ {[Myr]}} & 
\makecell{$P_{\text{cav}}$ \\ {[10$^{43}$ erg s$^{-1}$]}} & 
\makecell{$L_{\text{cool}}$ \\ {[10$^{43}$ erg s$^{-1}$]}} & 
\makecell{$t_{\text{cool}}$ \\ {[Gyr]}}\\
 \hline
 Abell 1068 (1)&$9.4\times 10^{-12}$ & $172.3 \times 10^{40}$&- &- & 0.36  & 14 & 0.35 $\pm$ 0.04 \\
\hline
Abell 1835 (2)&$14.7\times 10^{-12}$ & $163.9 \times 10^{40}$& 1.76  &40  & 140  & 120  & 0.30 $\pm$ 0.04 \\
\hline
Abell 2204 (3)&$21.9\times 10^{-12}$ & $159.4 \times 10^{40}$& 200  & 127  & 5000  & 160 & - \\
\hline
Abell 2390 (4)& $9.6\times 10^{-12}$ & $61.6 \times 10^{40}$& 468  &250  & 594  &40.4  & 0.64 $\pm$ 0.04 \\
\hline
RXJ1720+26& $14.3\times 10^{-12}$ & $12.7 \times 10^{40}$& -&- & - & -& 0.53 $\pm$ 0.13 \\
\hline
Abell 115&$9.0\times 10^{-12}$&$12.7 \times 10^{40}$& - &- & - & -&- \\
\hline
ZwCl 8276&$16.4\times 10^{-12}$ & $12.5 \times 10^{40}$& - &- & - & -& - \\
\hline
Abell 1795 (5)& $68.1\times 10^{-12}$ & $11.3 \times 10^{40}$& 0.1  &20  & 16  & 63  & 0.72 $\pm$ 0.09 \\
\hline
Abell 478 (6)&$39.9\times 10^{-12}$ & $10.8 \times 10^{40}$& 0.09  &30  & 10  & 144  & 0.46 $\pm$ 0.02 \\
 \hline
2A0336+096 (7)& $80.5\times 10^{-12}$ & $10.3 \times 10^{40}$& 0.04  &50  & 2.4  &33.8  & 0.31 $\pm$ 0.02\\
\hline
\textbf{Abell 2009}& $9.2\times 10^{-12}$ &$6.1 \times 10^{40}$& 0.63 & 20.2 & 112.4  &43.7 & 1.31 $\pm$ 0.09 \\
\hline
ZwCl 235 (8)& $10.9\times 10^{-12}$ & $4.1 \times 10^{40}$& 0.01  &17  & 1.2 & 10  & 0.75 $\pm$ 0.15\\
\hline
Abell 2199 (9)& $96.8\times 10^{-12}$ & $2.7 \times 10^{40}$& 0.20  &24  & 27  &14.2  & - \\
\hline
Abell 2052 (10)& $47.1\times 10^{-12}$ & $2.6 \times 10^{40}$& 0.01  &15  & 2.4  & 33.8  & 0.61 $\pm$ 0.10 \\
\hline
Abell 1668 (11)& $9.3\times 10^{-12}$ & $2.3 \times 10^{40}$& 0.001  &5.2  & 0.89  &1.9  & 1.40 $\pm$ 0.28 \\
\hline
Abell 2495 (12)& $11.8\times 10^{-12}$ & $2.0 \times 10^{40}$& 0.03  &17.9  & 4.7  &5.7  & 1.0 $\pm$ 0.2 \\
\hline
Abell 2634& $23.1\times 10^{-12}$ & $1.3 \times 10^{40}$& -& -& - &- & - \\
\hline
Abell 1991 (13)& $9.4\times 10^{-12}$ &$1.1 \times 10^{40}$& 0.06  &38.3  & 4.8  &6.04  & 0.41 $\pm$ 0.04 \\
\hline
\end{tabular}
\tablefoot{The X-ray fluxes and H$\alpha$ luminosities are taken from \citealt{1998MNRAS.301..881E} and \citealt{1999MNRAS.306..857C}, respectively. The references for the other quantities, from forth to seventh column, are: (1) \cite{2004ApJ...601..173M}, (2) \cite{2006ApJ...648..164M}, (3) \cite{2009MNRAS.393...71S}, (4) \cite{Sonkamble_2015}, (5) \cite{Russell_2017}, (6) \cite{2003ApJ...587..619S}, (7) \cite{2016ApJ...832..148V}, (8) \cite{2023AandA...670A..23U}, (9) \cite{2013ApJ...775..117N}, (10) \cite{2011ApJ...737...99B}, (11) \cite{2021ApJ...911...66P}, (12) \cite{2024ApJ...963....8R}, (13) \cite{Pandge_2013}. The reported cavity age, $t_{\text{cav}}$, is the average of the ages of the detected cavities, while the cavity energy and power values, $E_{\text{cav}}$ and $P_{\text{cav}}$, are the sum of the individual values for each cavity. In some cases, the properties of the identified cavities were not reported (Abell 115 ands ZwCl 8276), while in some other cases no cavities were detected in the existing exposures (RXJ1720+26 and Abell 2634). The quantities without a reported uncertainty are associated with a 50\% relative uncertainty in the plots of Fig.\ref{sample} (see Sec.\ref{sampleprop} for details). All the values for cooling time, t$_{cool}$, and their errors are taken from \cite{2008ApJ...687..899R}, except for Abell 2009 (from this work), ZwCl 235 (from \citealt{2023AandA...670A..23U}), Abell 1668 (from \citealt{2021ApJ...911...66P}), and Abell 2495 (from \citealt{2024ApJ...963....8R}).}
\label{samplevalues}
\end{table*}
The aim of the present work is to perform a detailed analysis of Abell 2009 by means of archival {\it Chandra} (20 ks) and JVLA (45 minutes) observations, in order to investigate its dynamical state and emission properties, and derive, for the first time, a comprehensive picture of the AGN feeding and feedback cycle within the systems. As the study of this cluster completes the dedicated investigation of the 18 objects in the BCS subsample (see Tab.\ref{samplevalues}), we also aim at providing a general overview of how heating and cooling proceeds in these representative examples of cool-core systems.\\
In this work, a $\Lambda$CDM cosmology is assumed, with H$_0$ = 70 km s$^{-1}$ Mpc$^{-1}$, $\Omega_0$ = 0.3, and $\Omega_{\Lambda}$ = 0.7. At the cluster redshift (z = 0.152), the angular size distance $D_{\rm A}$ is 545.8 Mpc, the luminosity distance $D_{\rm L}$ is 724.6 Mpc, and the scale corresponds to 2.65 kpc/$''$.


\begin{figure*}[ht!]
    \centering
    \begin{minipage}{0.52\textwidth}
        \centering
        \includegraphics[width=\textwidth]{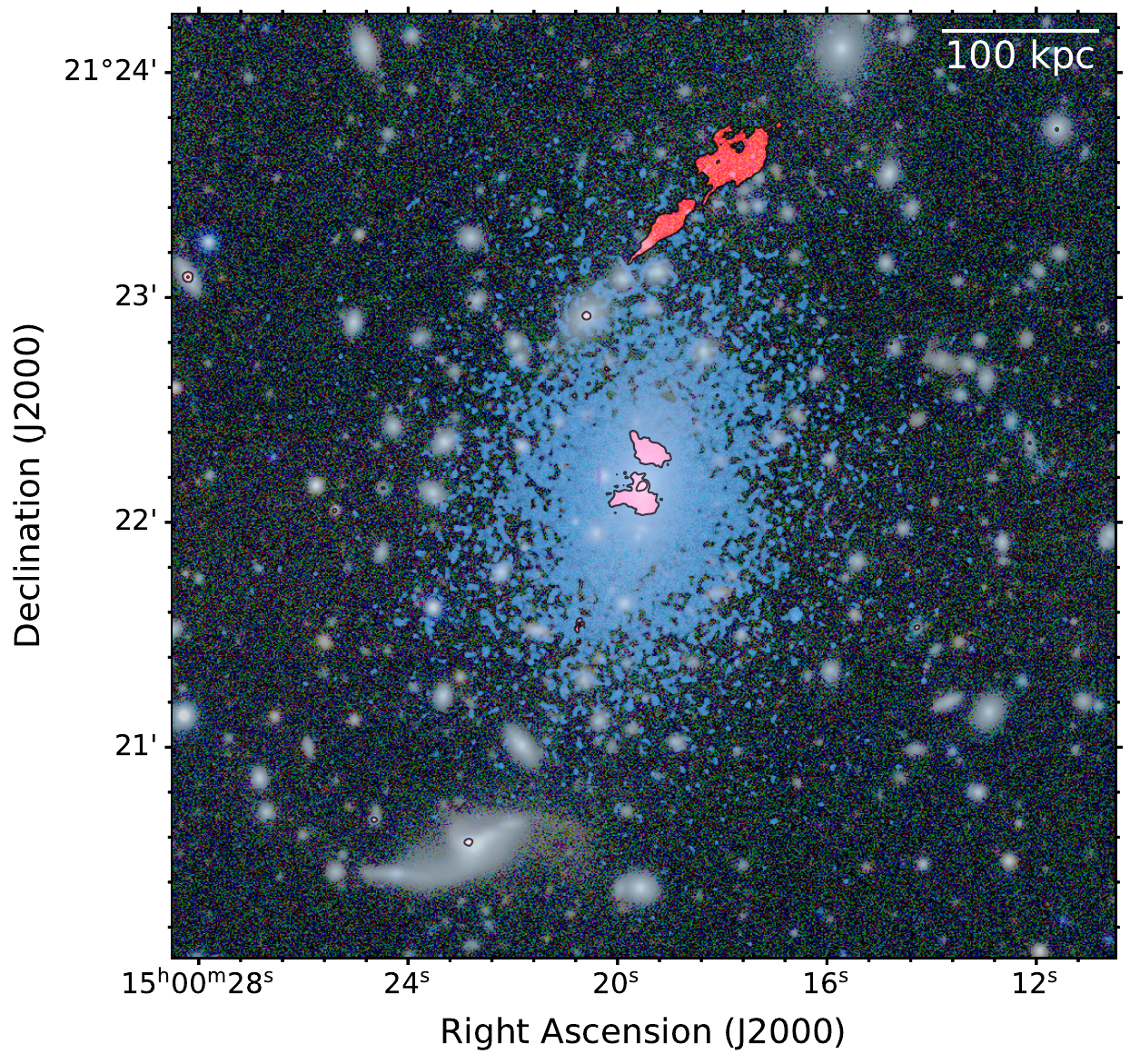}
    \end{minipage}
    \hspace{0.02\textwidth}
    \begin{minipage}{0.43\textwidth}
        \centering
        \includegraphics[width=\textwidth]{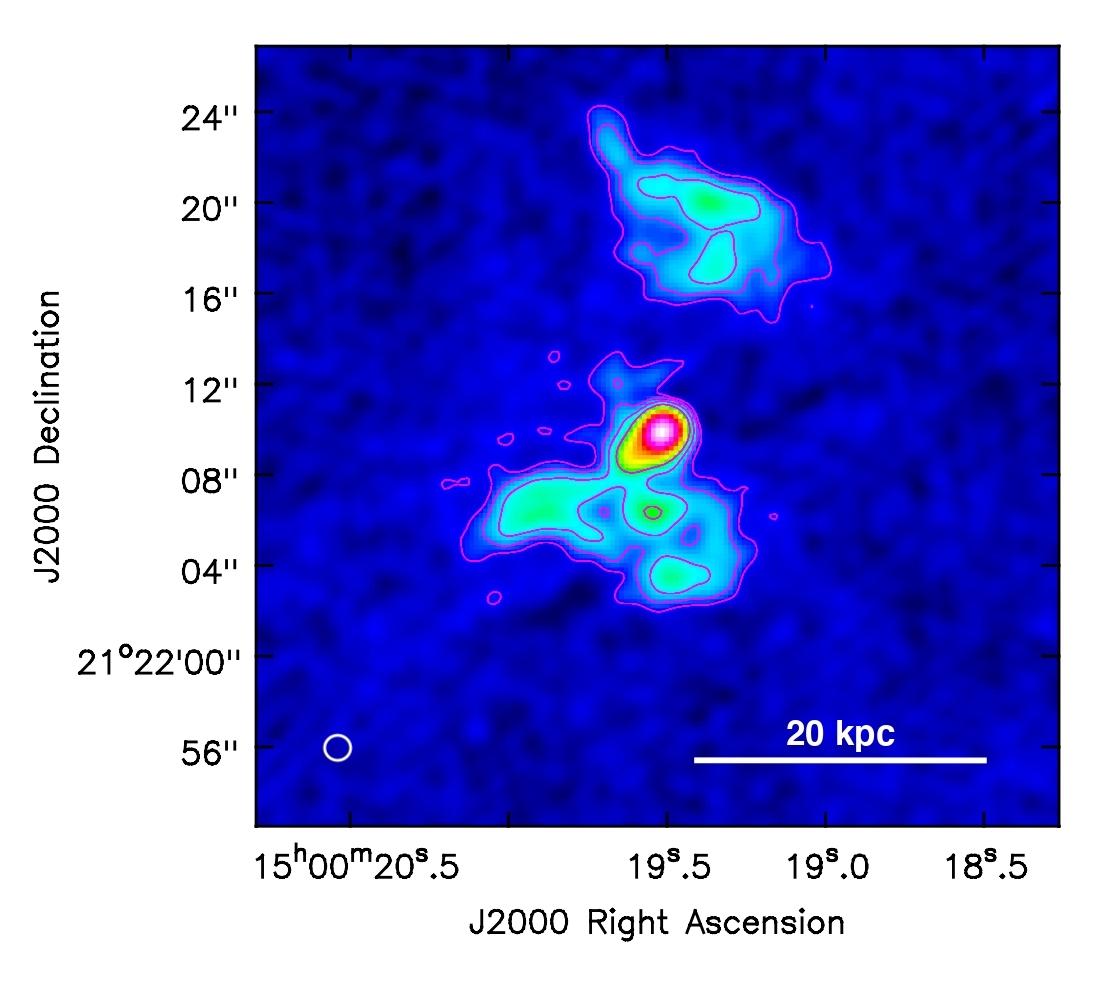} \\
        \vspace{4.8mm}
        \includegraphics[width=\textwidth]{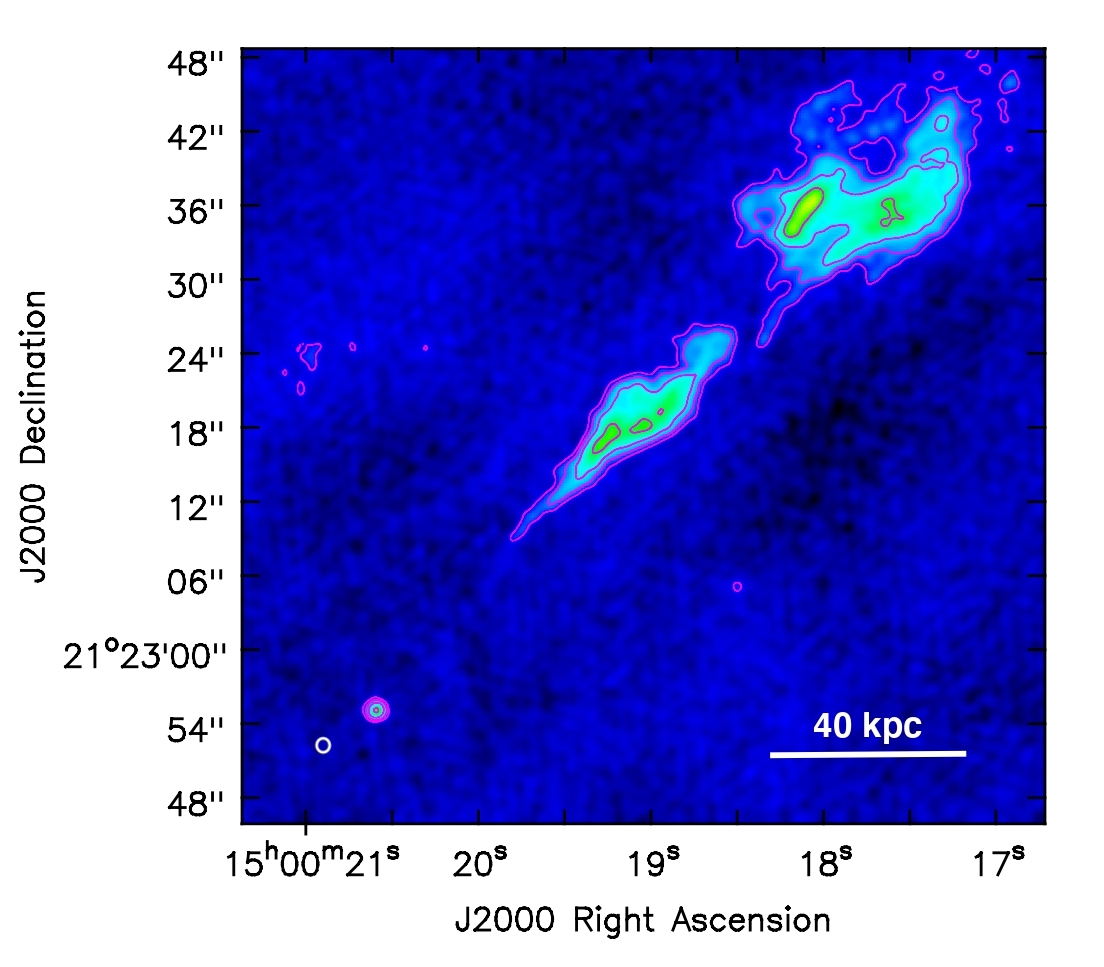}
    \end{minipage}
    \caption{Multiwavelength view of Abell~2009. Left panel: In this image, the radio emission at 1.5 GHz observed with the JVLA telescope and the X-ray emission detected by the \textit{Chandra} telescope are shown in red and blue, respectively. Both emissions are overlaid on an optical image from the DESI survey for a comprehensive view of the system. Upper right panel: 1.5 GHz JVLA radio image zoomed on the BCG, at a resolution of $1.2''\times 1.1''$. The rms value is $\sigma_{rms}$ = 19 $\mu$Jy/beam. Magenta contour levels are taken at 3, 6, 12, 24 $\times \sigma_{rms}$. Lower right panel: 1.5 GHz JVLA radio image zoomed on the northern radio galaxy, at a resolution of $1.2''\times 1.1''$. The rms value is $\sigma_{rms}$ = 19 $\mu$Jy/beam. Magenta contour levels are taken at 3, 6, 12, 24 $\times \sigma_{rms}$.}
    \label{radiogal}
\end{figure*}

\section{Data and methods}

\subsection{Chandra X-ray data}
Abell 2009 was observed with the \textit{Chandra} telescope during the Cycle 10 (ObsID 10438) for an exposure time of 19.9 ks. The data were acquired using ACIS-I instrument in VFAINT mode. The data reduction has been performed with the \texttt{CIAO(v4.15)} software (Chandra Interactive Analysis of Observations, \citealt{2006SPIE.6270E..1VF}) and the calibration data \texttt{CALDB(v4.10.2)}. In order to reprocess the data, the \textsf{chandra$\_$repro} script was used. The astrometry correction was performed by cross-matching the point sources (detected using the tool \textsf{wavdetect}) with two different catalogs: the United States Naval Observatory (USNO-A2.0) and the Gaia Data Release 2 (Gaia-DR2) optical catalogs. The best correction is achieved by using Gaia-DR2 catalog, owing to its high spatial resolution of $\sim$ 200 mas \citep{2018}, compared with the $\sim$ 250 mas resolution of USNO-A2.0 catalog. The final astrometric correction on the \textit{Chandra} dataset is of 1.3$''$ in the horizontal direction and -0.6$''$ in the vertical direction. The data were filtered to exclude time intervals with background flares. After running the \textsf{deflare} task, the cleaned exposure is 17.3 ks (87\% of the original exposure). The \textsf{blank-sky} script was used to automatically identify the appropriate blank-sky background file, which was subsequently normalized by the 9 - 12 keV count rate of the observation.\\
\\
In order to investigate the X-ray morphology of the ICM, the target, exposure and background images were produced in the 0.5 - 7 keV energy range. From these images, surface brightness radial profiles were extracted and fitted using \texttt{Sherpa}. 
The spectral analysis was performed using the \texttt{Xspec(v.12.13.0c)} software. The spectra were fitted in the energy range 0.5 - 7 keV and grouped with at least 25 counts in each spectral bin, enough to apply the $\chi^2$ statistic. Each spectrum is associated with ancillary response files (ARFs) and response matrix files (RMFs), and the blank-sky background spectrum is subtracted. The weight parameter is set to yes, which spatially weights the source ARF, as appropriate for extended sources. Moreover, the point sources are excluded from each region selected, in order to avoid contamination to the thermal spectra of the ICM. For every model used, we included a {\ttfamily{tbabs}} model component, representing galactic absorption due to the medium interposed between the source and the observer. The column density of galactic neutral hydrogen has been fixed to $N_H = 3.18\times10^{20}\text{cm}^{-2}$ \citep{2016A&A...594A.116H}. Moreover, we adopted the table of solar abundances from \cite{2009ARA&A..47..481A}.

\subsection{JVLA radio data}
To complement the X-ray analysis of Abell 2009, we considered radio observations that could provide radio maps with a resolution comparable to that of {\it Chandra} ($\lessapprox 1''$) and good sensitivity to extended emission from radio galaxies in the cluster. To this end, we consider the JVLA observations at L band (1-2 GHz) with the A configuration ($\sim 1''$ resolution), performed in 2014 (project 14A-280). The primary calibrator for this observation is 1331+305=3C286, while the secondary calibrator is J1513+2338. The target Abell 2009 was observed for 45 minutes. The data were calibrated using CASA v.6.4.1 \citep{casateam2022} following standard data reduction procedures for continuum observations\footnote{See \url{https://casaguides.nrao.edu/}.}. We attempted self-calibration of the target, but its relatively low peak flux density (about 6 mJy at 1.5 GHz) prevented us from obtaining useful phase and amplitude solutions. In any case, the rms noise of our images (about 20 $\mu$Jy/beam) is close to the theoretical expected one (about 15 $\mu$Jy/beam). Images were obtained in CASA with the task \texttt{tclean} adopting multiscale cleaning and multifrequency synthesis. We found that imaging with \texttt{robust = 0.5} offers the best compromise between spatial resolution and sensitivity to extended emission. Uncertainties on the measured flux densities were obtained using the equation
\begin{equation}
    \Delta S = \sqrt{(\sigma_{c} \cdot S)^{2} + (RMS \cdot \sqrt{N_{beam}})^{2}},
\end{equation}
where $S$ is the flux density, $N_{beam}$ is the number of beams in the source area, and $\sigma_{c}$ is the systematic calibration error on the flux, which is typically assumed of 5\% for JVLA \citep{1998AJ....115.1693C}.


\section{Results}
In this section, we outline the results of the morphological and spectral analysis of Abell 2009.

\subsection{Radio morphological analysis}\label{radio}
In Fig.\ref{radiogal} (left panel), we present the 1.5 GHz JVLA radio emission (in red) and the \textit{Chandra} X-ray emission (in blue) of Abell 2009, overlaid on an optical image for a multiwavelength view of the system. The image reveals the radio emission of two different sources: the radio galaxy associated with the BCG and another radio galaxy in the north direction.
The BCG emission presents a bright core and extended radio lobes ($\sim$ 17 kpc long), oriented from northwest (NW) to southeast (SE), which are symmetrically distributed in a butterfly-shaped morphology (see Fig.\ref{radiogal} upper right panel). The arched and concave morphology of the lobes may be caused by precession of the jets around their axis, by projection effects, or both (e.g., \citealt{2008MNRAS.384.1327S}, \citealt{2008ApJ...682..155W}, \citealt{2013ApJ...777..163H}). We measured that this radio galaxy has a total flux density of 23 $\pm$ 1.0 mJy above 3$\sigma$ level, with 5.8 $\pm$ 0.3 mJy coming from the central bright core (see Tab.\ref{fluxes}). The radio powers at 1.5 GHz are calculated assuming a spectral index\footnote{$S_{\nu}\propto \nu^{\alpha}$} $\alpha =-1$ for the BCG (\citealt{2008AJ....136..684K}) and $\alpha =-0.3$ for the core component (\citealt{2015MNRAS.453.1201H}), using the formula:
\begin{equation}
    P_{radio}=4\pi D_L^2 S_{\nu} (1+z)^{-(\alpha +1)}.
\end{equation}
Their estimates are $(1.45\pm 0.06)\times 10^{24}$ W Hz$^{-1}$ and $(0.33\pm 0.02)\times 10^{24}$ W Hz$^{-1}$, respectively (see Tab.\ref{fluxes}). Considering its morphology and the radio power value, this galaxy can be associated to an FRI type (see \citealt{1996AJ....112....9L}).\\
The radio galaxy in the N direction, located approximately at 50$''$ from the BCG, presents an elongated radio lobe and, along the direction of the lobe toward SE, a compact core can be observed (see Fig.\ref{radiogal} lower right panel), which corresponds to an elliptical galaxy at the same redshift of the cluster (2MASX J15002058+2122548, z = 0.157 from \citealt{2020ApJS..249....3A}). Thus, this radio galaxy is a cluster member of Abell 2009 and has a total flux density of 41.0 $\pm$ 1.6 mJy (including the core emission), corresponding to a 1.5 GHz radio power of $(2.77\pm 0.11)\times 10^{24}$ W Hz$^{-1}$, calculated assuming $\alpha =-1$ (see Tab.\ref{fluxes}).

\begin{table*}[h]
    \centering
    \caption{Properties of the radio sources in Abell 2009.}
    \begin{tabular}{cccccc}
    \hline
    \noalign{\smallskip}
     & RA& dec & z & F$_{1.5 GHz}$ [mJy] & P$_{1.5 GHz}$ [10$^{24}$ W/Hz]\\
    \noalign{\smallskip}
    \hline
    \noalign{\smallskip}
     \makecell{BCG (2MASX J15001950+2122108) \\ core}& 15:00:19.5 & +21:22:09.9 & 0.152& \makecell{23.0 $\pm$ 1.0 \\ 5.8 $\pm$ 0.3} & \makecell{ 1.45 $\pm$ 0.06 \\ 0.33 $\pm$ 0.02}\\
    \noalign{\smallskip}
    \hline
    \noalign{\smallskip}
     N galaxy (2MASX J15002058+2122548)& 15:00:20.6 & +21:22:55.1 & 0.157& 41.0 $\pm$ 1.6 & 2.77 $\pm$ 0.11\\
    \noalign{\smallskip}
    \hline
\end{tabular}
    \tablefoot{This table reports the coordinates, redshift, radio flux densities [mJy], and radio powers [10$^{24}$ W/Hz] at 1.5 GHz of each source.}
    \label{fluxes}
\end{table*}

\subsection{X-ray morphological analysis}
In this section, we describe the X-ray surface brightness profile extraction and the fit with both a single $\beta$ model and a double $\beta$ model. Subsequently, we describe the 2D fitting analysis, deriving the residual image used to investigate the presence of ICM substructures.

\subsubsection{Surface brightness profile}
\begin{figure}[h]
   \centering
  \includegraphics[width=0.40\textwidth]{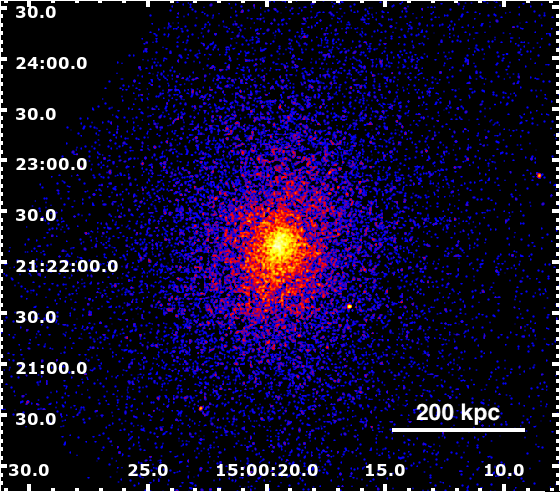}
   \caption{0.5 - 7 keV \textit{Chandra} image of the galaxy cluster Abell~2009, convolved with a Gaussian function of a 2 pixel radius.}
   \label{abell2009}
\end{figure}

In Fig.\ref{abell2009}, we show the 0.5 - 7 keV \textit{Chandra} image of Abell 2009. The azimuthally averaged surface brightness profile is extracted with \texttt{CIAO}. 
The extraction regions are concentric circular annuli centered on the BCG coordinates, which is located at the X-ray peak at the center of the cluster. The annuli extend to a maximum radius of 131$''$ ($\sim$ 347 kpc) and are spaced with a 1$''$ width, except for the outermost three that are wider ($\approx$ 4$''$) to include more counts. The emission from the X-ray point sources across the field was excluded, as well as the innermost 2 bins of the profile, where contamination from the AGN in the BCG may be present. 
\begin{figure}[!h]
        \centering
        \includegraphics[width=.40\textwidth]{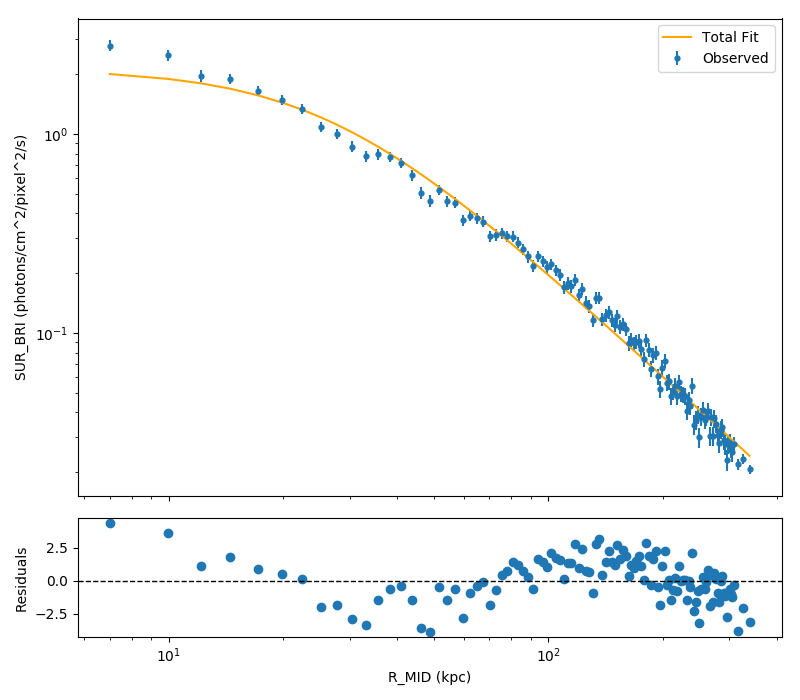}
        \caption{0.5-7 keV surface brightness profile of Abell~2009 fitted with a single $\beta$ model. The lower subplot represents the residuals in $\sigma$ units.}
        \label{fig.3.3}
\end{figure}\\
Initially, the surface brightness profile is fitted with a single $\beta$ model \citep{1978A&A....70..677C} defined as:
\begin{equation}
    S(r)= A \left[ 1+\left(\frac{r}{r_0}\right)^2 \right]^{0.5-3\beta},
    \label{sbetamodel}
\end{equation}
where $A$ is the amplitude of the model, $r_0$ is the core radius, and $\beta$ is the parameter that controls the slope of the profile at large radii. The best-fit parameters are reported in Tab.\ref{tab11} and the fitted profile in Fig.\ref{fig.3.3}. As it can be seen from the profile, the model does not fit accurately the surface brightness in the central regions, where there is an evident excess of emission. This discrepancy is a characteristic feature of cool-core clusters, where the denser cores undergo radiative cooling and exhibit elevated X-ray luminosity (e.g., \citealt{Lewis_2003}).
Moreover, the reduced $\chi^2$ value is 322.9/115 = 2.8, meaning that the model cannot properly represent the data.\\
To properly describe the surface brightness profile, we fitted it with a double $\beta$ model, defined as:
\begin{equation}
     S(r)= A_1 \left[ 1+\left(\frac{r}{r_{0,1}}\right)^2 \right]^{0.5-3\beta_1} + A_2 \left[ 1+\left(\frac{r}{r_{0,2}}\right)^2 \right]^{0.5-3\beta_2},
\end{equation}\\
\\
where $A_1$ and $A_2$ are the amplitudes, $r_{0,1}$ and $r_{0,2}$ are the core radii, and $\beta_1$ and $\beta_2$ are the parameters that controls the slopes of each component.
The fit is performed both by considering different values for the two $\beta$ \citep{2000MNRAS.318.1041E} and by linking them to the same value \citep{1999ApJ...517..627M}. The best-fit parameters of both cases are reported in Tab.\ref{tab11} and their respective fitted profiles in Fig.\ref{fig.3.4}. 
\begin{table*}[h!]
\centering
\caption{Single and double $\beta$ model fits.}
\begin{tabular}{ccccccccc}
\hline
\noalign{\smallskip}
  Model& $\beta_1$& $\beta_2$ &r$_{0,1}$ {\fontsize{10}{12}\selectfont [kpc]}& r$_{0,2}$ {\fontsize{10}{12}\selectfont [kpc]} &A$_1$ &A$_2$ & $\chi^2$/dof\\
  \noalign{\smallskip}
 \hline
 \noalign{\smallskip}
  1$\beta$& 0.46 $\pm$ 0.01& -& 26.71 $\pm$ 0.98 & - & 2.13 $\pm$ 0.08& -& 322.9/115 (2.81)\\
  \noalign{\smallskip}
  \hline
 \noalign{\smallskip}
  2$\beta$& 0.59 $\pm$ 0.05& 0.57 $\pm$ 0.19& 104.3 $\pm$ 23.8 & 18.4 $\pm$ 6.3 & 0.38 $\pm$ 0.20& 2.77 $\pm$ 0.42& 126.2/112 (1.13)\\
  \noalign{\smallskip}
  \hline
  \noalign{\smallskip}
  2$\beta$ (linked)& 0.58 $\pm$ 0.02& = $\beta_1$ & 102.9 $\pm$ 8.1 & 18.8 $\pm$ 1.6 & 0.40 $\pm$ 0.03 & 2.74 $\pm$ 0.17 & 126.2/113 (1.12) \\
  \noalign{\smallskip}
\hline
\end{tabular}
\tablefoot{Parameters values for the single and double $\beta$ model fits. The first row represents the single $\beta$ model fit, the second row corresponds to the fit with the two $\beta$ values unconnected, while the third row represents the fit with the linked $\beta$ values. The first two columns report the values of $\beta$, the third and forth columns report the core radii in [kpc] units, the fifth and sixth columns report the amplitudes of the models in [10$^{-7}$ photons cm$^{-2}$ pixel$^{-2}$ s$^{-1}$] units, and the last column corresponds to the reduced $\chi^2$ values.}
\label{tab11}
\end{table*}
\begin{figure*}[h!]
  \centering
  \includegraphics[width=0.40\textwidth]{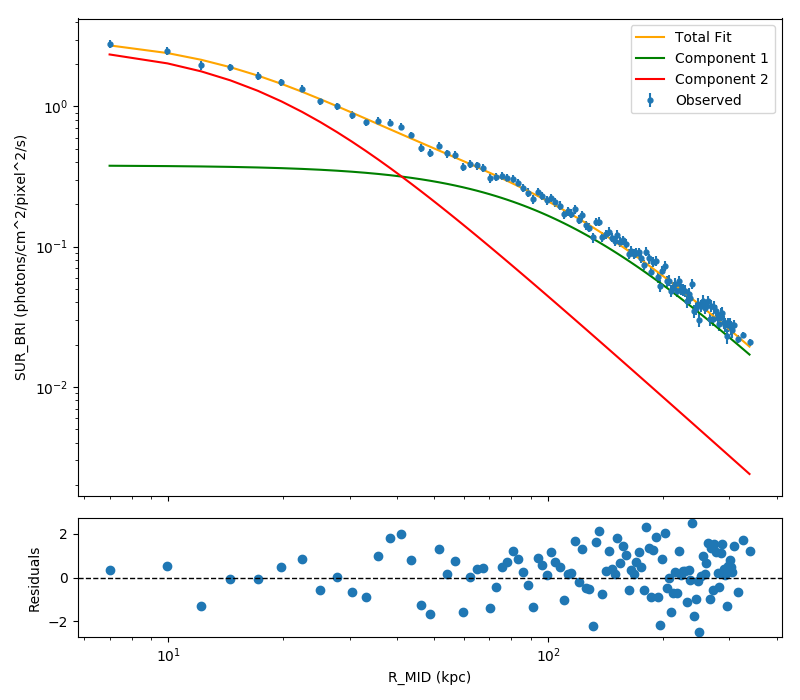}
  \hspace{0.05\textwidth}
  \includegraphics[width=0.40\textwidth]{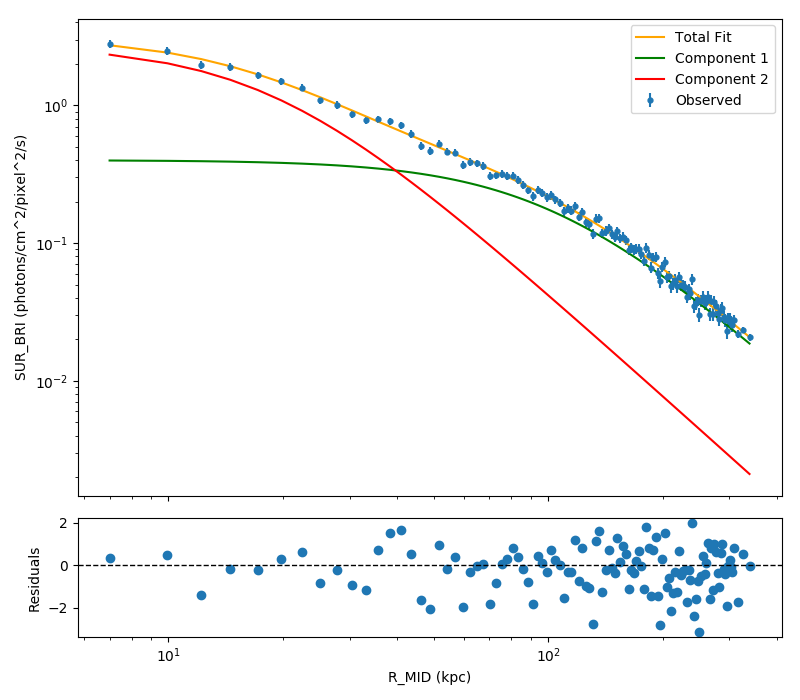}
  \caption{Left panel: 0.5-7 keV surface brightness profile of Abell 2009 fitted with a double $\beta$ model where the values of $\beta$ are unlinked (see Tab.\ref{tab11}). The orange line represents the total model, while the red and green lines correspond to the two components of the model. The lower subplot represents the residuals in $\sigma$ units. Right panel: 0.5-7 keV surface brightness profile of Abell 2009 fitted with a double $\beta$ model where the values of $\beta$ are linked (see Tab.\ref{tab11}). The orange line represents the total model, while the red and green lines correspond to the two components of the model. The lower subplot represents the residuals in $\sigma$ units.}
  \label{fig.3.4}
\end{figure*}\\
From the reduced $\chi^2$ values in Tab.\ref{tab11}, it is clear that this model better fits the data compared to the single $\beta$ model. The fit in which the two $\beta$ values were linked (see Tab. \ref{tab11}, last row) yields best-fit parameters and statistics that are consistent with the case of different $\beta$ values (see also e.g., \citealt{2024ApJ...963....8R}). For this reason, it is reasonable to consider the model with the linked $\beta$ values as the best fit to the data.

\subsubsection{Searching for X-ray cavities}\label{cav}
\begin{figure}[h!]
   \centering
  \includegraphics[width=0.40\textwidth]{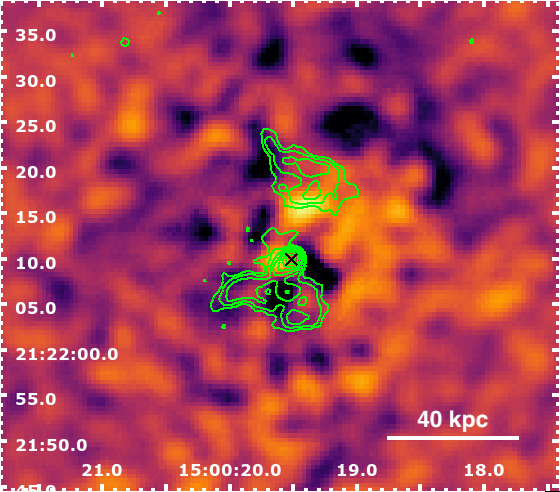}
   \caption{Residual image of the best-fit double $\beta$ model convolved with a Gaussian function of a 5 pixel radius. The green contours identify the radio emission at 1.5 GHz. The rms noise is $\sigma_{rms}$ = 19 $\mu$Jy/beam. The contour levels start from 3$\sigma$ and increase by a factor 2, while the beam size is 1.2$'' \times$ 1.1$''$. The black cross corresponds to the position of the BCG.}
   \label{resid+contours}
\end{figure}

In order to investigate the possible presence of substructures in the ICM, we derived a residual (model subtracted) image of the X-ray emission from Abell 2009. Specifically, the 0.5-7 keV image of Abell 2009, subtracted from the background and corrected by the exposure, is fitted with a 2D double $\beta$ model in \texttt{Sherpa}. The extraction region corresponds to the largest possible circle centered in the BCG coordinates with a radius of 131$''$ ($\sim$347 kpc).
The free parameters of the fit are the core radius, $r_0$, the ellipticity of the model, $ellip$ ($\epsilon=1-\frac{b}{a}$, where $a$ is the semi-major axis and $b$ is the semi-minor axis of the ellipse), the angle of the major axis, $theta$, the amplitude of the model, $ampl$, and the power-law slope of the profile at large radii, $\delta$. The slope $\beta$ of the single power-law model directly relates to the brightness profile, while the slope $\delta$ of the double power-law corresponds to the exponent (0.5 - 3$\beta$) of Eq.\ref{sbetamodel} and by equating the two expressions, the following relationship is obtained: $\beta=(\delta+0.5)/3$.
The best-fit parameters are reported in Tab.\ref{tab13}. 
\renewcommand{\arraystretch}{1.7}
\begin{table*}[h!]
\centering
\caption{2D double $\beta$ model fit.}
\begin{tabular}{ccccccc}
\hline
  $\delta$& r$_0$ {\fontsize{10}{12}\selectfont [kpc]}& A {\fontsize{10}{12}\selectfont [10$^{-7}$photons cm$^{-2}$pixel$^{-2}$s$^{-1}$]} &ellip & theta & $\chi^2$/dof {\fontsize{10}{12}\selectfont [10$^{-15}$]}\\
\hline
   \shortstack{\\ 0.91$^{+0.01}_{-0.01}$ \\ 2.62$^{+0.02}_{-0.02}$} & \shortstack{\\ 43.05$^{+2.33}_{-2.32}$ \\ 20.19$^{+1.25}_{-1.24}$} & \shortstack{\\ 3.12$^{+0.02}_{-0.02}$ \\ 5.32$^{+0.03}_{-0.03}$} & \shortstack{\\ 0.27$^{+0.01}_{-0.01}$ \\ 0.27$^{+0.01}_{-0.01}$} & \shortstack{\\ 1.35$^{+0.02}_{-0.02}$ \\ 1.35$^{+0.02}_{-0.02}$} & \multicolumn{1}{c}{\raisebox{1.4ex}{1.32}} \\
\hline
\end{tabular}
\tablefoot{Parameters values for the 2D double $\beta$ model fitting the exposure corrected image. The upper values are the one for the first component and the lower ones are for the second component. The first column reports the value of $\delta$, the second column reports the core radius in [kpc] units, and the third column reports the amplitude of the model in [10$^{-7}$photons cm$^{-2}$ pixel$^{-2}$ s$^{-1}$] units. The forth and fifth columns report the values for ellipticity and angle of inclination, and the last column reports the reduced $\chi^2$ value (it results relatively small due to the image being scaled by the exposure map and the number of degrees of freedom being high). All the errors reported are determined with a 1$\sigma$ confidence level.}
\label{tab13}
\end{table*}\\
In Fig.\ref{resid+contours}, we show the residual image derived by subtracting the 2D fit from the original \textit{Chandra} image, with the radio contours at 1.5 GHz superimposed. The comparison between the residual map and the radio contours shows no clear evidence of depressions in the ICM that align with the radio lobes. We further evaluated the significance of several negative residual patches visible in the image, but none were found to be statistically significant. However, we note that this does not necessarily imply the absence of cavities. Given the short exposure time of the \textit{Chandra} observations ($\sim 17$ ks) , potential cavities may not be detectable at the current sensitivity level. 
In order to evaluate any feedback effect from this radio galaxy on the ICM, in the following we consider the radio lobe volumes as tracers of the mechanical energy injected by the AGN jets, as previously done in the literature in other cases where cavities were not detected in X-rays (e.g., \citealt{2022A&A...668A..65T}, \citealt{2022MNRAS.513.3273S}).\\
We also searched for possible surface brightness edges in Abell 2009 that could trace cold fronts or shocks. However, we could not confirm the presence of any significant density jump in the existing {\it Chandra} observations. While it is possible that the cluster does not currently host cold or shock fronts, it is also possible that the relatively short {\it Chandra} exposure of 20~ks is not deep enough to reveal shallow and localized gradients in surface brightness.

\subsection{X-ray spectral analysis}
In this section, we describe the global spectral extraction performed to obtain the large-scale properties of the cluster Abell~2009. Subsequently, we outline the methods to perform a thorough projected and deprojected analysis. The latter is essential to provide an effective representation of the thermodynamical radial profiles of the ICM and, consequently, to understand the dynamical state of the gas.

\subsubsection{Global spectral properties of Abell 2009}
To derive a global spectrum of the ICM and measure the basic thermodynamic properties, such as temperature and metallicity, we considered an ellipse centered on the BCG with a semi-major axis of 164$''$ ($\sim$434 kpc), a semi-minor axis of 130$''$ ($\sim$343 kpc), and an inclination angle of 77.3$^{\circ}$. This extraction region corresponds to the largest area that can be covered within the {\it Chandra} field of view. The regions of the point sources are excluded from the global spectral analysis. The fit of the spectrum is performed using a  {\ttfamily{tbabs*apec}} model, which includes the component of galactic absorption of the medium interposed between the source and the observer (tbabs), and the one of thermal emission (apec). The second component accounts for the thermal emission from a collisional ionized gas, with parameters of temperature kT, metal abundance Z, redshift z, and normalization norm of the spectrum. The spectrum is then fitted with this model and the results are shown in Tab.~\ref{global}.
\begin{table*}[h!]
\centering
\caption{Global spectrum best-fit of Abell 2009.}
\begin{tabular}{cccccc}
\hline
  Net counts&kT [keV] & Z [Z$_{\odot}$]& norm [10$^{-2}$]&$\chi^2$/d.o.f.  \\
 \hline
  23270 (96\%)& 6.41$^{+0.17}_{-0.17}$ & 0.59$^{+0.07}_{-0.07}$ & 1.20$^{+0.02}_{-0.02}$ &311.68/311 (1.00)\\
\hline
\end{tabular}
\tablefoot{Best-fit results of the global spectrum of Abell 2009. In the first column, the net photon counts in the 0.5 - 7 keV energy band and their percentage with respect to the total counts are reported. The second, third, and forth columns are for the temperature, metallicity, and normalization. The last column reports the reduced $\chi^2$ value. }
\label{global}
\end{table*}
The best-fit parameters are kT = 6.41 $\pm$ 0.17 keV and Z = 0.59 $\pm$ 0.07 Z$_{\odot}$, which are consistent with the typical values of galaxy clusters (\citealt{bahcall1996clusters}).

\subsubsection{Projected radial analysis}\label{proj}
\begin{figure}[ht]
  \centering
  \includegraphics[width=0.46\textwidth]{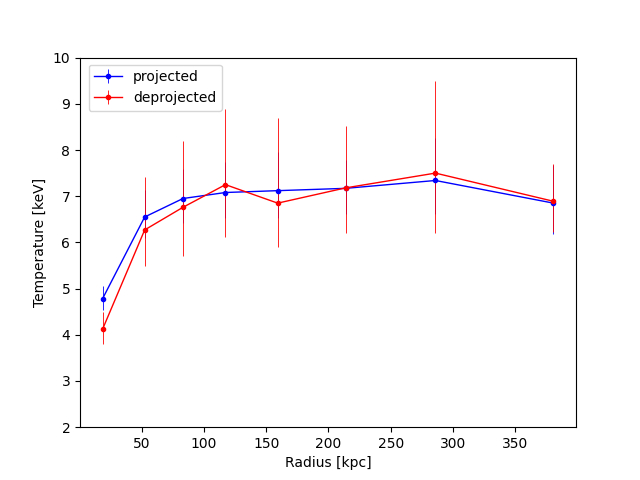}
  \caption{Projected (blue) and deprojected (red) radial temperature profiles of Abell 2009. The temperature measurements are reported in Tab.\ref{projected} and Tab.\ref{deprojected}.}
  \label{kTprof}
\end{figure}

This analysis is performed to derive the profiles of the projected thermodynamical quantities. The spectra are extracted from eight elliptical and concentric annuli, centered in the coordinates of the BCG. The spacing between the annuli is determined by the requirement that inside each annulus there must be at least 2000 net counts, in order to accurately constrain the fit parameters. The annuli are chosen with the same ellipticity and inclination angle of the surface brightness distribution of the cluster (see Subsec.\ref{cav}). The maximum extension is an ellipse with a semi-major axis of 164$''$ ($\sim$434 kpc) and a semi-minor axis of 130$''$ ($\sim$343 kpc). Furthermore, the central 1.5$''$ ($\sim$4 kpc) have not been considered, since any contribution from the X-ray emission of the AGN may contaminate the ICM spectrum in that region.
Each annulus spectrum is fitted with a {\ttfamily{tbabs*apec}} model and the results are shown in Tab.\ref{projected}.
In Fig.\ref{kTprof}, we show the projected temperature profile. As it can be observed, the temperature decreases going toward small radii, which is typical of cool-core clusters \citep{Peterson_2006}.

\subsubsection{Deprojected radial analysis}\label{deproj}
In the projected analysis, the derived quantities suffer from the contribution of the ICM from the outer regions that is projected along the line of sight and the normalization parameter cannot be used to calculate the electron density in each annulus. The deprojection analysis is performed in order to obtain radial quantities without the influence of the projection effect. The same spectra used for the projected analysis were loaded on \texttt{Xspec} and fitted with a {\ttfamily{projct*tbabs*apec}} model. The {\ttfamily{projct}} mixing model component performs a 3D to 2D projection of prolate ellipsoidal shells onto elliptical annuli.  The best-fit results are listed in Tab.\ref{deprojected} and, in Fig.\ref{kTprof}, we show the deprojected temperature profile superimposed to the corresponding projected profile. 
\setlength{\tabcolsep}{3pt}
\begin{table*}[h!]
\centering
\caption{Best-fit parameters of the deprojected radial spectral analysis of Abell 2009.}
\begin{tabular}{ccccccccc}
\hline
  R$_{in}$ - R$_{out}$ [kpc]&kT [keV]& Z [Z$_{\odot}$]& norm [10$^{-3}$] &$\chi^2$/d.o.f.& $n_e$ [10$^{-2}$ cm$^{-3}$] & p [10$^{-1}$ keV cm$^{-3}$]& K [keV cm$^2$]  \\
 \hline
  3.9 - 37.4& 4.12$^{+0.37}_{-0.32}$ & 1.07 (fixed) & 1.25$^{+0.04}_{-0.04}$ & 66.76/86 (0.78) & 4.22 $\pm$ 0.07 & 3.18 $\pm$ 0.26 & 33.99 $\pm$ 2.64 \\
\hline
37.4 - 67.6& 6.27$^{+1.15}_{-0.79}$ & 0.80$^{+0.48}_{-0.43}$& 1.31$^{+0.11}_{-0.11}$ & 84.21/76 (1.11) & 1.94 $\pm$ 0.08& 2.23 $\pm$ 0.31& 86.84 $\pm$ 10.96\\
\hline
67.6 - 98.9& 6.76$^{+1.43}_{-1.06}$ & 0.59 (fixed) & 1.44$^{+0.08}_{-0.08}$  & 79.19/75 (1.06) & 1.28 $\pm$ 0.11& 1.58 $\pm$ 0.26& 123.54 $\pm$ 19.39\\
\hline
98.9 - 135.8& 7.25$^{+1.63}_{-1.14}$ & 0.96 (fixed) & 1.53$^{+0.08}_{-0.08}$  & 66.44/79 (0.84) & 0.86 $\pm$ 0.02& 1.14 $\pm$ 0.19& 172.72 $\pm$ 27.19 \\
\hline
135.8 - 182.7& 6.85$^{+1.84}_{-0.94}$ & 0.28 (fixed)& 2.05$^{+0.09}_{-0.10}$  & 82.05/83 (0.99) & 0.65 $\pm$ 0.01& 0.82 $\pm$ 0.12& 196.67 $\pm$ 27.04 \\
\hline
182.7 - 244.7& 7.18$^{+1.33}_{-0.97}$ & 0.78 (fixed)& 2.13$^{+0.09}_{-0.09}$  & 84.40/83(1.02) & 0.43 $\pm$ 0.01& 0.57 $\pm$ 0.09& 271.11 $\pm$ 36.72 \\
\hline
244.7 - 326.9& 7.50$^{+1.99}_{-1.29}$ & 0.48$^{+0.49}_{-0.46}$ & 1.97$^{+0.16}_{-0.16}$  & 74.60/79 (0.94) & 0.27 $\pm$ 0.01& 0.37 $\pm$ 0.08& 387.76 $\pm$ 67.26\\
\hline
326.9 - 434.2& 6.89$^{+0.80}_{-0.66}$ & 0.37$^{+0.28}_{-0.26}$ & 3.14$^{+0.15}_{-0.15}$  & 76.00/77 (0.99) & 0.22 $\pm$ 0.01& 0.28 $\pm$ 0.04& 403.66 $\pm$ 39.40\\
\hline
\end{tabular}
\tablefoot{Best-fit results of the elliptical deprojected radial spectral analysis of Abell~2009. In the first column the internal and outer radii of each elliptical annulus are reported (these radii correspond to the semi-major axes of the annuli). The second, third, forth, and fifth columns report the temperature, metallicity, normalization, and reduced $\chi^2$ values of the fit, while the sixth, seventh, and eighth columns report the electron density, pressure, and entropy values.}
\label{deprojected}
\end{table*}

\begin{table*}[h!]
\centering
\caption{Best-fit results of the spectral analysis of the cooling region in Abell 2009.}
\begin{tabular}{ccccccc}
\hline
  R$_{in}$ - R$_{out}$ [kpc]& kT [keV] & Z [Z$_{\odot}$]& norm [10$^{-2}$]& Net counts &$\chi^2$/d.o.f.   \\
 \hline
  3.9 - 87.5& 5.56 $^{+0.26}_{-0.24}$  & 0.77 $^{+0.14}_{-0.13}$ & 0.46 $^{+0.01}_{-0.01}$ & 7124 (97.7\%) & 145.65/164 (0.89)\\
\hline
87.5 - 447.8& 7.01 $^{+0.29}_{-0.26}$& 0.50 $^{+0.09}_{-0.09}$& 1.06 $^{+0.02}_{-0.02}$ & 13588 (85.7\%) & 222.55/251 (0.89)\\
\hline
\end{tabular}
\tablefoot{Best-fit results of the elliptical deprojected radial spectrum of Abell 2009 using a {\ttfamily{projct*tbabs*apec}} model and derived to study the inner cooling region. In the first column the internal and outer radii of each elliptical annulus are reported (these radii correspond to the semi-major axes of the annuli). The second, third, and forth columns report the temperature, metallicity, and normalization values. The last two columns report the net photon counts and their percentage with respect to the total counts, and the value for the reduced $\chi^2$ of each fit.}
\label{minflow1}
\end{table*}
The deprojected profile is broadly consistent with the projected one, showing a central decrease in temperature and a flat profile at $r\gtrapprox100$~kpc. The central temperature drop appear slightly more pronounced than in the projected profile, although the larger uncertainties associated with the more complex deprojection model prevent any meaningful comparison.
Using the normalization values of the deprojected analysis, we calculated the electron density of the plasma for each annulus. The normalization is defined as:
\begin{equation}
    norm=\frac{10^{-14}}{4\pi [D_A(1+z)]^2} \int n_e n_H dV,
\end{equation}
where $D_A$ is the angular distance from the source (D$_A$ = 545.8 Mpc for Abell 2009), z is the redshift, $n_e$ and $n_H$ are the electron and proton densities, and V is the volume of the emitting region. It is possible to invert this equation and derive the electron density:
\begin{equation}
    n_e=\sqrt{10^{14}\left( \frac{4\pi \times norm \times [D_A (1+z)]^2}{0.83V} \right)}.
\end{equation}
The electron density values for each annulus are reported in Tab.\ref{deprojected} and its profile in Fig.\ref{thermo} (upper left panel).
Subsequently, the pressure and entropy profiles were calculated from the density profile, assuming $n_e\approx 1.2 n_p$ and using the following formulae: $p=1.83 \, n_ekT$ and $K=kT/n_e^{2/3}$. The best-fit values of p and K are shown in Tab.\ref{deprojected} and their profiles in Fig.\ref{thermo} (upper right and lower left panels). The thermodynamical profiles exhibit characteristic features of cool-core clusters \citep{Peterson_2006}. Specifically, there is a clear increase in density toward the cluster center, while temperature and entropy decrease. The derivation of these thermodynamic properties is essential for computing other key quantities, such as the cooling time.

\subsubsection{Cooling properties of Abell 2009}\label{cooling}
\begin{figure*}[h]
    \centering
    \begin{subfigure}[b]{0.245\textwidth}
        \centering
        \includegraphics[width=\textwidth,trim={0.2cm 0.2cm 0.2cm 0.2cm},clip]{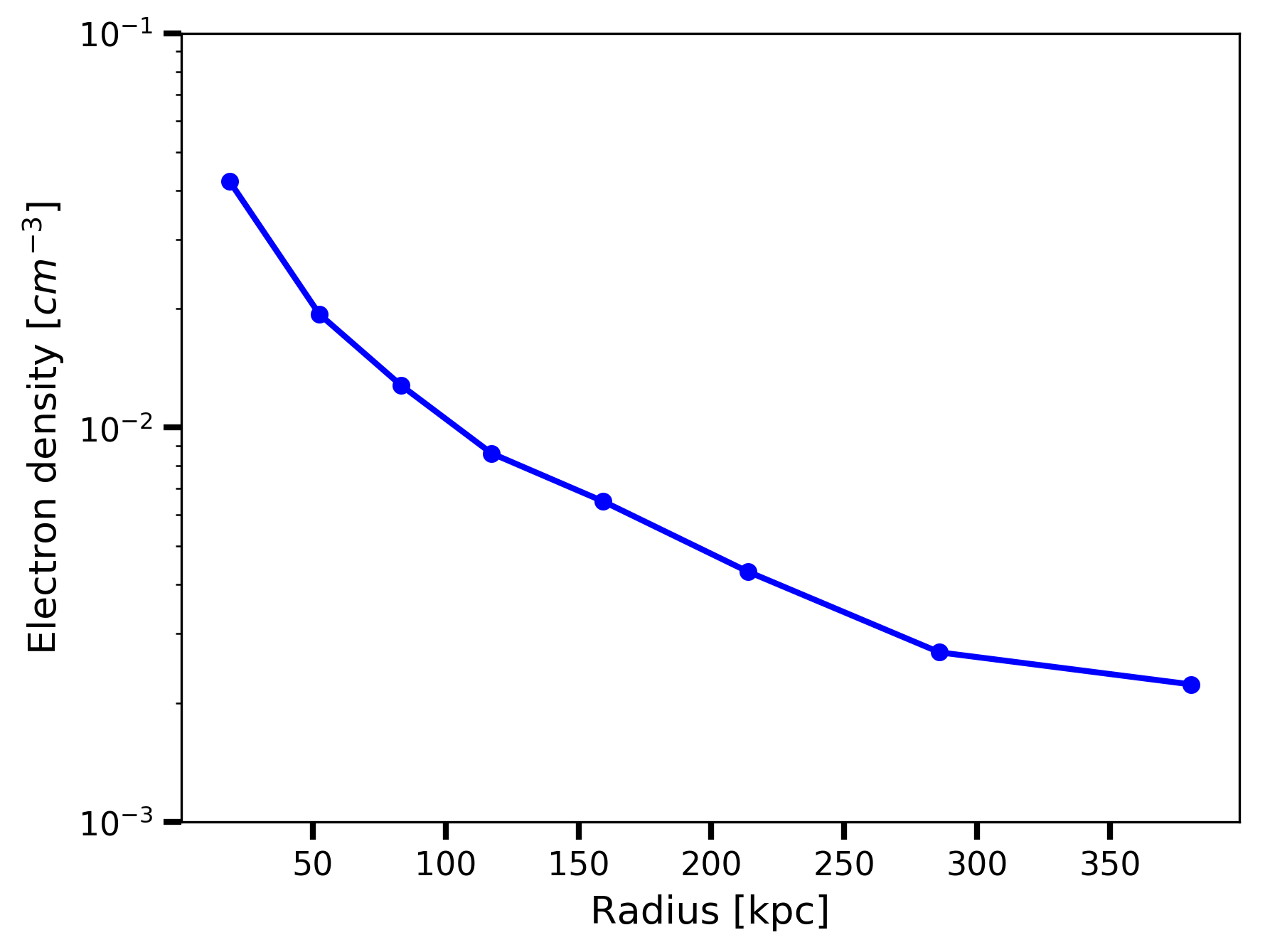}
        \label{}
    \end{subfigure}
    \begin{subfigure}[b]{0.245\textwidth}
        \centering
        \includegraphics[width=\textwidth,trim={0.2cm 0.2cm 0.2cm 0.2cm},clip]{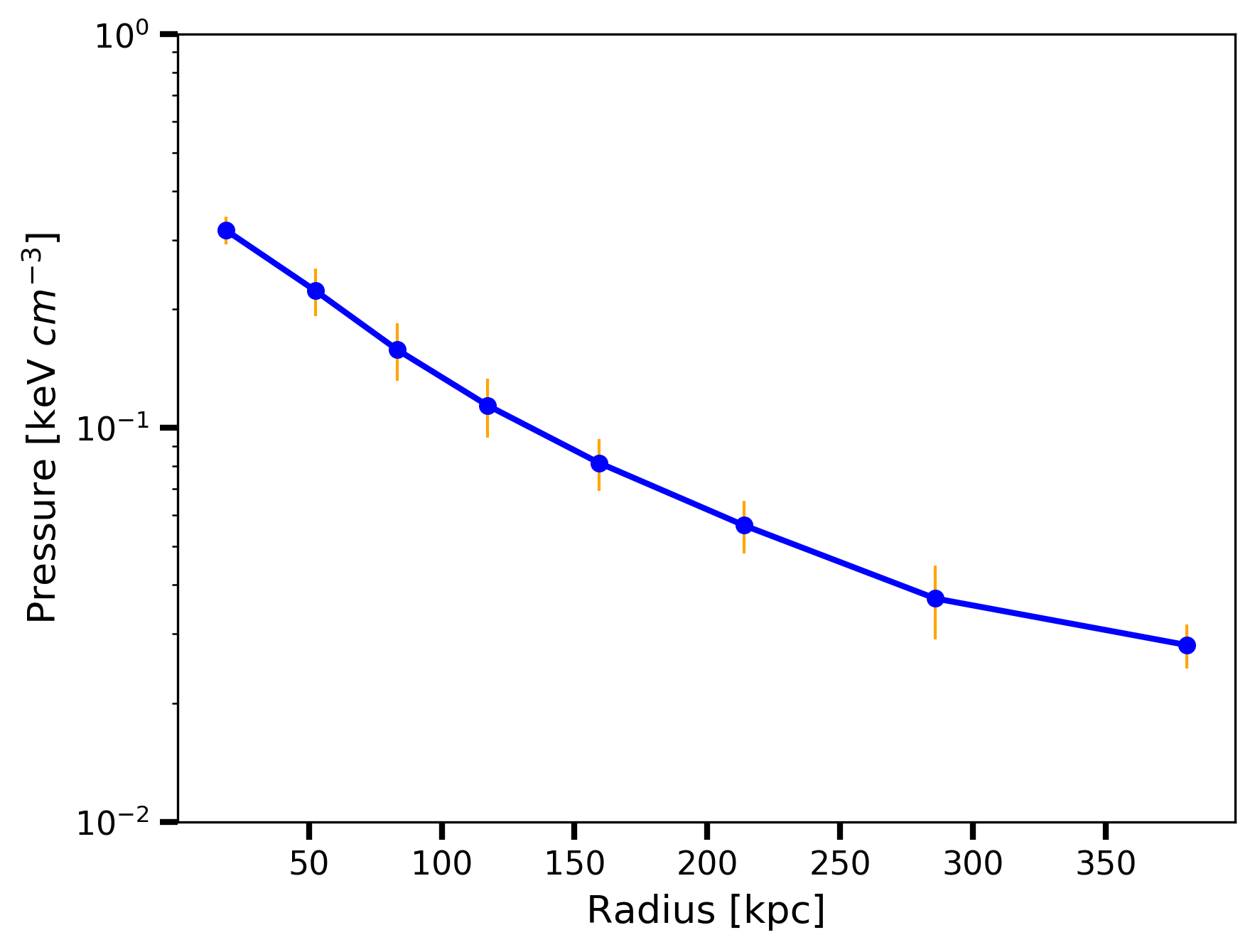} 
        \label{}
    \end{subfigure}
    \begin{subfigure}[b]{0.245\textwidth}
        \centering
        \includegraphics[width=\textwidth,trim={0.2cm 0.2cm 0.2cm 0.2cm},clip]{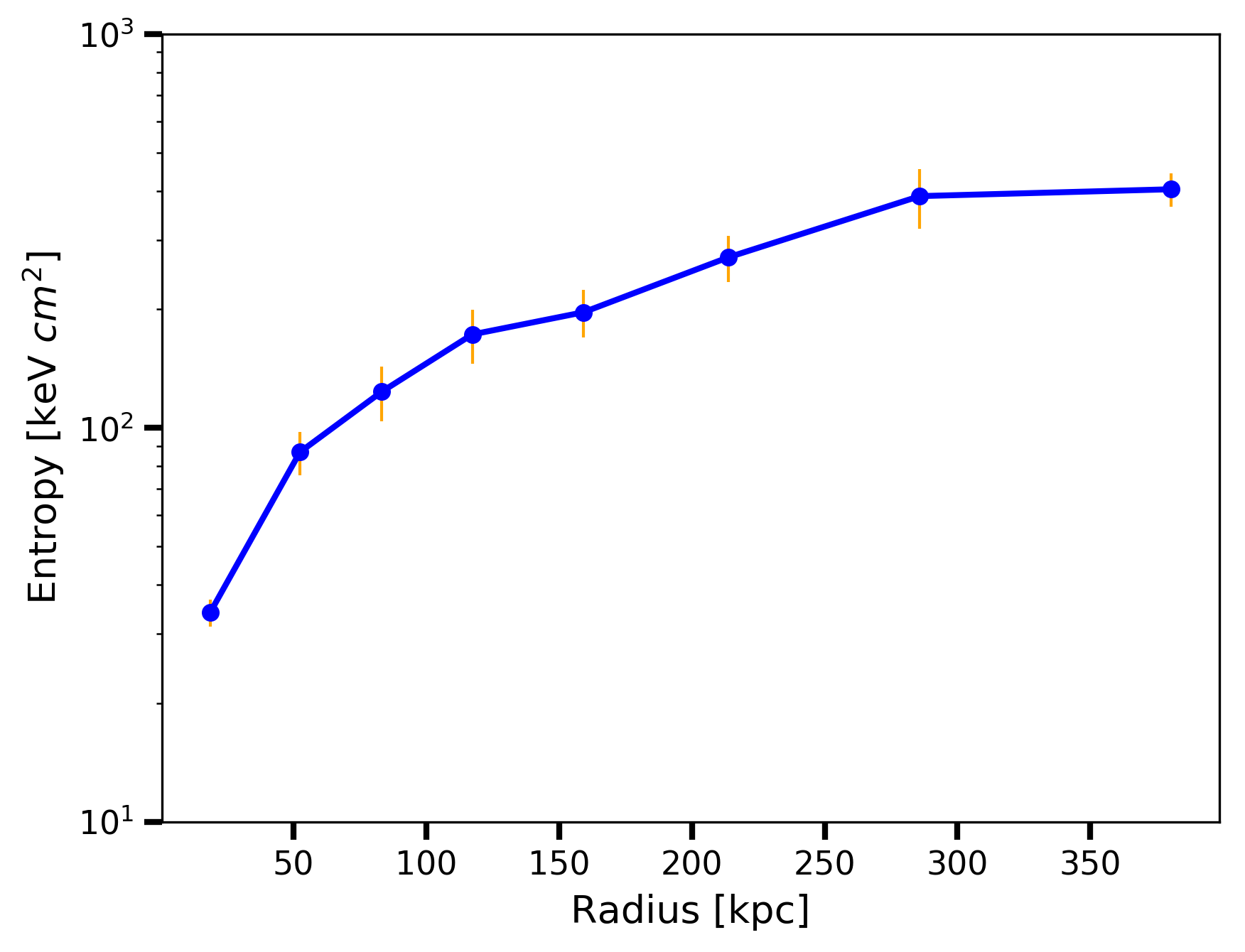} 
        \label{}
    \end{subfigure}
    \begin{subfigure}[b]{0.245\textwidth}
        \centering
        \includegraphics[width=\textwidth,trim={0.2cm 0.2cm 0.2cm 0.2cm},clip]{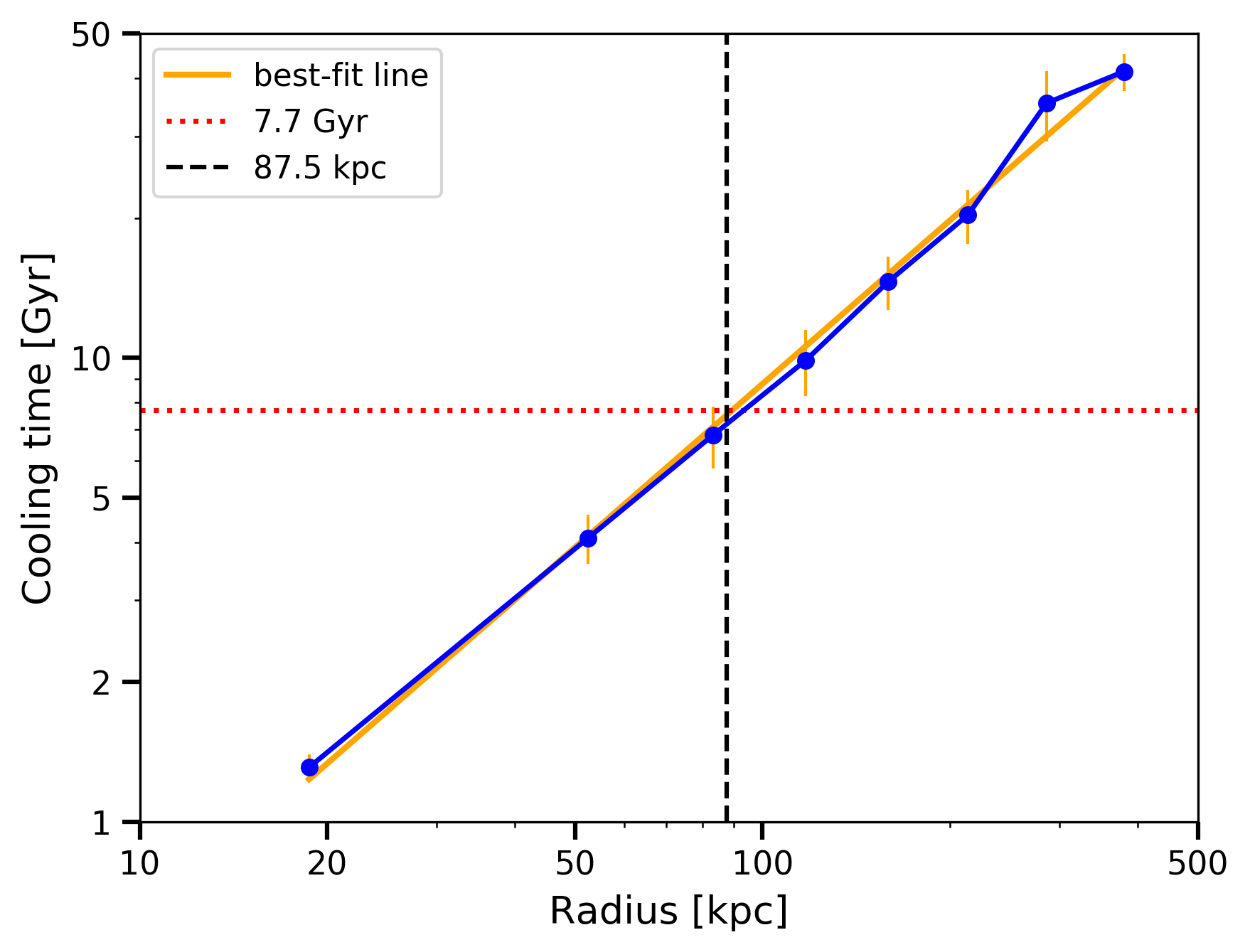} 
        \label{}
    \end{subfigure}
    
    \caption{Radial profiles of thermodynamic properties of Abell 2009. First panel: Deprojected electron density profile of Abell 2009. Second panel: Deprojected pressure profile of Abell 2009. Third panel: Deprojected entropy profile of Abell 2009. Forth panel: Cooling time profile fitted with a polynomial fit (yellow line) showing the interception with the red horizontal line corresponding to a 7.7 Gyr cooling time and the black vertical line corresponding to the cooling radius value (87.5 kpc).}
    \label{thermo}
\end{figure*}

Since the surface brightness analysis revealed a profile typical of a relaxed cluster, with a central peak in brightness, and the thermodynamic profiles align with expectations for a cool-core, we now move on to analyze the cooling properties of the cluster. 
In order to identify the cooling region, we calculate the cooling time profile considering the temperature and density profiles, and the value of the cooling function $\Lambda (T)$ at each annulus, using the equation (e.g., \citealt{Peterson_2006}, \citealt{Gitti_2012}):
\begin{equation}
    t_{cool}=\frac{\gamma}{\gamma -1} \frac{kT}{\mu X n_e \Lambda(T)} \label{tcool},
\end{equation}
where $\gamma$ is the adiabatic index, $\mu$ is the molecular weight, $X$ is the hydrogen mass fraction, and $\Lambda(T)$ is the cooling function (see \citealt{1993ApJS...88..253S}) .
The cooling function is calculated with the use of \texttt{CLOUDY} code, which simulates fully physical conditions within an astronomical plasma and predicts the emitted spectrum \citep{1998PASP..110..761F}. We show the cooling time profile in Fig.\ref{thermo} (lower right panel), which presents lower values at smaller radii, as expected for cool-core clusters. The cooling radius, $r_{cool}$, is usually defined as the radius at which $t_{cool}$ is equal to the look-back time at z=1 ($\sim$ 7.7 Gyr; e.g., \citealt{2004ApJ...607..800B}, \citealt{McNamara_2007}, \citealt{Gitti_2012}). In order to accurately determine its value for Abell~2009, we consider the intersection between the reference value of 7.7 Gyr (represented by a horizontal line in Fig.\ref{thermo} lower right panel) and a polynomial fit to the cooling time profile, finding a cooling radius of:
\begin{equation}
    r_{cool}=87.5 \pm 15.7 \text{ kpc},
\end{equation}
where the uncertainty is given by the extension of the radial bin that includes the cooling radius value.
In order to study the properties of the cooling region, a spectral analysis of the emission inside the cooling radius is performed. The spectra are extracted from two elliptical annuli centered in the coordinates of the central AGN. The first annulus extends from a semi-major axis of 1.5$''$ ($\sim$ 4 kpc) to 33.1$''$ ($\sim$ 87.5 kpc, which corresponds to the cooling radius). The maximum extension of the second annulus is given by a semi-major axis of 169.2$''$ ($\sim$ 448 kpc).
First, the spectra are fitted with a {\ttfamily{projct*tbabs*apec}} model, where the redshift z and the column density $N_H$ were fixed, and the best-fit values are reported in Tab.\ref{minflow1}.
We then measured the bolometric (0.1 - 100 keV) X-ray luminosity, or the cooling luminosity, of the cooling region, finding:
\begin{equation}
    L_{cool}=(4.4\pm 0.1) \times 10^{44} \text{ erg s$^{-1}$}. \label{lcool}
\end{equation}
Using this luminosity value, the predicted global mass inflow rate of gas in the cooling region is calculated using the equation (see \citealt{1994ARA&A..32..277F}):
\begin{equation}
    L_{cool}=\frac{5}{2} \frac{kT}{\mu m_p} \dot{M}\label{L-Mdot},
\end{equation}
and the result is $\dot{M}_{glob}= (316 \pm 16)$ M$_{\odot}$/yr.\\
Finally, the spectra are fitted with a {\ttfamily{projct*tbabs*(apec+mkcflow)}} model to account for the amount of cooling gas. Here, the additive component {\ttfamily{mkcflow}} describes the cooling flow model of \cite{1988ASIC..229...53M}. The spectral analysis returned only an upper limit on the spectral mass inflow rate, of $\dot{M}_{spec} \leq 19.4$ M$_{\odot}$/yr. In Tab.\ref{minflow2}, all the properties derived for the cooling region are reported.
\begin{table*}[h!]
\centering
\caption{Characteristics of the cooling region of Abell 2009.}
\begin{tabular}{cccccc}
\hline
  R$_{in}$ - R$_{out}$ [kpc]& L$_{cool}$ [10$^{44}$ erg s$^{-1}$] & $\dot{M}_{glob}$ [M$_{\odot}$/yr]& $\dot{M}_{spec}$ [M$_{\odot}$/yr] & $\chi^2$/dof\\
 \hline
  3.9 - 87.5& 4.37 $\pm$ 0.09& 316 $\pm$ 16& $\leq$ 19.4& 145.65/163 (0.89)\\
\hline
\end{tabular}
\tablefoot{In the first column the internal and outer radii of the cooling region are reported (these radii correspond to the semi-major axes of the annuli). The subsequent three columns report the cooling luminosity (see Eq.\ref{lcool}), global mass inflow rate (derived using Eq.\ref{L-Mdot}), and spectral mass inflow rate (derived usign a {\ttfamily{projct*tbabs*(apec+mkcflow)}} model) values. The last column reports the reduced $\chi^2$ value of the fit using the {\ttfamily{projct*tbabs*(apec+mkcflow)}} model.}
\label{minflow2}
\end{table*}
The fitting model including the {\ttfamily{mkcflow}} component does not represent a significant improvement compared to that considering only the thermal {\ttfamily{apec}} component, since the reduced $\chi^2$ assumes the same value of 0.89. 
Moreover, comparing the global and spectral mass inflow rates, it can be clearly seen that the global value largely overestimates the spectral upper limit.

\section{Discussion}
In this Section, we first describe the AGN feedback impact on Abell 2009 and then discuss some general properties of the BCS subsample from which it was selected (see Tab.\ref{samplevalues}).

\subsection{AGN feedback in Abell 2009}
\begin{figure}[h!]
   \centering
  \includegraphics[width=0.40\textwidth]{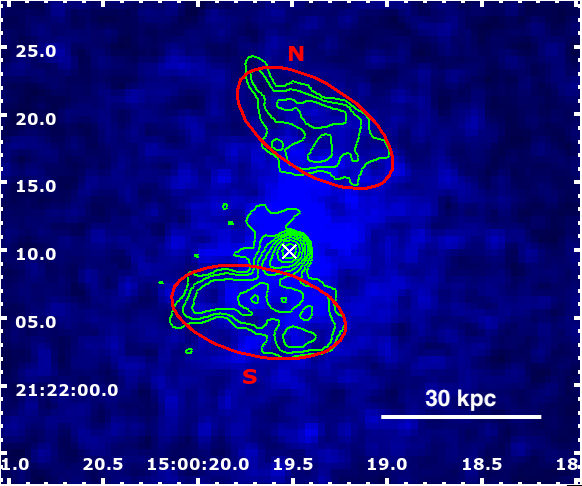}
   \caption{0.5-7 keV \textit{Chandra} image of Abell~2009 convolved with a Gaussian function of a 2 pixel radius. The green contours identify the radio emission at 1.5 GHz. The rms noise is $\sigma_{rms}$ = 19 $\mu$Jy/beam. The contour levels start from 3$\sigma$ and increase by a factor 2, while the beam size is 1.2$''$x1.1$''$. The red regions correspond to the north and south regions used to describe the size of the radio lobes (N and S). The white cross corresponds to the position of the BCG.}
   \label{lobes}
\end{figure}

The spectral analysis revealed a cooling region with an extension of $r_{cool}=87.5\pm 15.7$ kpc, where the cooling time is lower than 7.7 Gyr. The global mass inflow rate obtained from the cooling luminosity of the cooling region is $\dot{M}_{glob}= (316 \pm 16)$ M$_{\odot}$/yr, which is derived considering the standard cooling flow model. If the cooling rate is, instead, spectroscopically determined considering the {\ttfamily{mkcflow}} component, which represents an isobaric multiphase cooling model, we obtain an upper limit $\dot{M}_{spec} \leq 19$ M$_{\odot}$/yr.
In order to understand the difference between the expected and observed mass inflow rates (presented in Subsec.\ref{cooling}), we wish to evaluate the efficiency of AGN feedback in heating the ICM and reducing the cooling losses within the cool core.
\\Given its projected location at about 130 kpc from the center of Abell 2009, the northern radio galaxy (described in Sec.\ref{radio}) is positioned outside the cooling radius. Thus, while we focus primarily on the central radio galaxy, the external radio galaxy may still contribute to heating processes, potentially impacting regions outside the core through its interaction with the ICM (e.g., \citealt{2022MNRAS.513.3273S}).
Since the central AGN is located within the cooling region, we assess its mechanical power to investigate the heating and cooling balance within the cluster core. Specifically, we consider the mechanical power of the radio lobes, assuming that these correspond to X-ray cavities (e.g., \citealt{2022A&A...668A..65T}, \citealt{2022MNRAS.513.3273S}), and compare it with the radiative losses within the cooling region, represented by the bolometric X-ray luminosity of the ICM (Eq. \ref{lcool}). 
We consider the 1.5 GHz extension of the radio lobes as representative of the size of any underlying cavity, and, assuming a prolate ellipsoid 3D shape, the volume is calculated as $V=\frac{4}{3}\pi ab^2$, where $a$ is the semi-major axis and $b$ is the semi-minor axis. In Fig.\ref{lobes}, the two regions describing the lobes can be visualized and, in Tab.\ref{cavities}, we report their morphological characteristics.
\begin{table}[h!]
\centering
\caption{Morphology of the BCG radio lobes of Abell 2009.}
\begin{tabular}{ccccc}
\hline
  &a [kpc]&b [kpc] & V [kpc$^3$]& d [kpc] \\
 \hline
   N lobe &17.2 $\pm$ 1.7& 8.6 $\pm$ 0.9& 5304 $\pm$ 749& 25.6 $\pm$ 2.6\\
\hline
   S lobe &17.2 $\pm$ 1.7& 8.6 $\pm$ 0.9& 5304 $\pm$ 749 & 12.9 $\pm$ 1.3\\
\hline
\end{tabular}
\tablefoot{The table reports the morphological characteristics of the radio lobes of the BCG. In the first two columns, the semi-major and semi-minor axes are reported. The third column reports the lobes volume. The last column reports the distance of the lobes from the central AGN.}
\label{cavities}
\end{table}\\
In order to derive the mechanical AGN power, $P_{cav}=4pV/t_{cav}$, we first calculate the pV work done by the AGN to inflate the lobe. We consider the ICM pressure within the first annulus of the deprojected spectral analysis (see Tab. \ref{deprojected}), which encompasses both radio lobes, measuring a mechanical energy of about  4pV = 3 $\times$ 10$^{59}$ erg for each lobe. 
The radio lobes seem to be detected at an early evolutionary state since they are relatively close to the AGN, at about 15 - 20 kpc from the center.  Thus, we compute the dynamical age of the radio lobes using the sound crossing time, $t_{sc}$. This timescale assumes that the bubbles are rising through the hot gas atmosphere at the ICM sound speed, defined as:
\begin{equation}
    c_s =\sqrt{\frac{\gamma kT}{\mu m_p}}.
\end{equation}
The sound crossing time of the lobes is thus: $t_{cav}=d/c_s$, where d is the distance of the lobe from the BCG (reported in Tab.\ref{cavities}). The sound speed in the first annulus is $c_s=936\pm 72$ km s$^{-1}$ and the corresponding lobes ages are approximately 26.8 Myr (N) and 13.5 Myr (S). The estimated mechanical AGN power values are thus $\sim$ 3.8 $\times$ 10$^{44}$ erg s$^{-1}$ (N) and 7.5 $\times$ 10$^{44}$ erg s$^{-1}$ (S). These results are reported in Tab.\ref{energetics}.
The power of the N and S lobes is summed to obtain the total cavity power: 
\begin{equation}
P_{cav}= (11.2 \pm 1.9) \times 10^{44} \text{ erg s$^{-1}$}, \label{pcav}
\end{equation}
where the error is calculated as the dispersion at 1$\sigma$ from the mean value.
\begin{table}[h!]
\centering
\caption{Energetics of the BCG radio lobes of Abell 2009.}
\begin{tabular}{cccc}
\hline
  &E$_{cav}$ [10$^{59}$ erg]&t$_{cav}$ [Myr] & P$_{cav}$ [10$^{44}$ erg s$^{-1}$] \\
 \hline
   N&3.17 $\pm$ 0.52& 26.76 $\pm$ 3.39& 3.76 $\pm$ 0.77 \\
\hline
   S&3.18 $\pm$ 0.52& 13.48 $\pm$ 1.71& 7.48 $\pm$ 1.53 \\
\hline
\end{tabular}
\tablefoot{The table reports the energy, age, and power values for the north and south radio lobes, obtained in the assumption that these trace X-ray cavities.}
\label{energetics}
\end{table}\\
We are aware that these estimates are relatively uncertain and that the real uncertainties are probably dominated by systematics, as the lobes could be older than measured with the sound crossing time. Considering other methods to derive the age of cavities and bubbles, the discrepancy could be up to a factor of two (e.g., \citealt{2004ApJ...607..800B}, \citealt{2021ApJ...923L..25U}). Projection effects constitute another important source of uncertainty. Additionally, the assumption of the radio lobes being representative tracers of cavities adds further uncertainties. On the one hand, any X-ray cavity could be smaller than the lobes at 1.5 GHz, thus leading to an overestimate of their volume (and thus the energy). However, the radio lobes may be larger at lower frequencies ($\approx$ 100s MHz), where radiative losses of relativistic electrons are less effective in dimming the synchrotron emission than at 1.5 GHz (see the example of MS~0735.6+7421; e.g., \citealt{2021A&A...650A.170B}). These two opposite effects prevent us from deriving an accurate estimate of the size of the cavities from the radio tracer. All in all, we consider that systematic uncertainties on energy and age of the radio lobes may be up to a factor of two, and those on the mechanical power may be up to a factor of four. 
The value of the X-ray luminosity inside the cooling region, obtained in Subsec.\ref{cooling}, is L$_{cool}$ = (4.4 $\pm$ 0.1) $\times$ 10$^{44}$ erg s$^{-1}$. This measure is a factor of two lower than the mechanical power estimated from the central radio galaxy's lobes, suggesting that the cooling luminosity (L$_{cool}$) can be balanced by the mechanical power (P$_{cav}$). Even if the cavities were assumed to be half as large and twice as old — reducing the mechanical power by a factor of four — the resulting power would still be approximately 3 $\times$ 10$^{44}$ erg s$^{-1}$, remaining comparable to the cooling luminosity. This indicates that, even under conservative assumptions, AGN mechanical feedback is likely sufficient to counterbalance radiative losses in Abell~2009’s core region.

\subsection{Clusters sample properties}\label{sampleprop}
Abell~2009 has been extracted from a ROSAT BCS subsample in which 18 clusters with X-ray fluxes greater than $4.4 \times 10^{-12}$ erg cm$^{-2}$ s$^{-1}$ \citep{1998MNRAS.301..881E} and $H\alpha$ luminosities greater than $10^{40}$ erg s$^{-1}$ \citep{1999MNRAS.306..857C} were selected (see Tab.\ref{samplevalues}).
To obtain a general perspective on the AGN feedback process, it is interesting to investigate how the ICM properties may vary with the central AGN ones in the objects of this subsample. We thus examine both the AGN properties, such as cavity power (P$_{cav}$), cavity energy (E$_{cav}$) and cavity age (t$_{cav}$), as well as the properties of the ICM, in particular the cooling luminosity (L$_{cool}$). For most objects, these measurements are available from the literature (see references in Tab.\ref{samplevalues})\footnote{In some cases, the properties of the identified cavities were not reported in literature, while in other cases no cavities were detected in the existing exposures (see caption of Tab.\ref{samplevalues}). When cavities are absent,
cooling may still be balanced through alternative mechanisms (as discussed in e.g., \citealt{2006ApJ...638..659M}, \citealt{McNamara_2007}). However, these non-detections should not significantly impact the correlations discussed in this work, as cavities are identified in
85\% of the cases within our selection.}. Given the nonhomogeneous analysis of the subsample, we assume a conservative 50\% relative uncertainty. Moreover, we also consider the cooling time at 12 kpc from the center, reported by \cite{2008ApJ...687..899R} for 9/18 of the clusters in our selection, along with the associated uncertainty, and we derive a similar estimate for Abell~2009 (from this work), ZwCl 235 (from \citealt{2023AandA...670A..23U}), Abell 1668 (from \citealt{2021ApJ...911...66P}), and Abell 2495 (from \citealt{2024ApJ...963....8R}). Such a small radius of 12 kpc should capture the most rapidly cooling part of the ICM. We note that, for Abell 2009 and Abell 2495, the innermost radial bin of the thermodynamic profile (and thus the smallest resolution element) is at a radius of 16.8 kpc and 23.6 kpc, respectively. Therefore, there may be slight variations with respect to the actual cooling time at 12 kpc. 
These measures are reported in Tab.\ref{samplevalues}.\\
We caution that the following discussion aims at exploring general trends between relevant quantities for our selection of systems and it is not intended to represent a statistical assessment of scaling relations for galaxy clusters in general.

\subsubsection{Cavity power versus cooling luminosity}\label{Pc-Lc}
\begin{figure*}[ht!]
    \centering
    \begin{subfigure}[b]{0.44\textwidth}
        \centering
        \includegraphics[width=\textwidth]{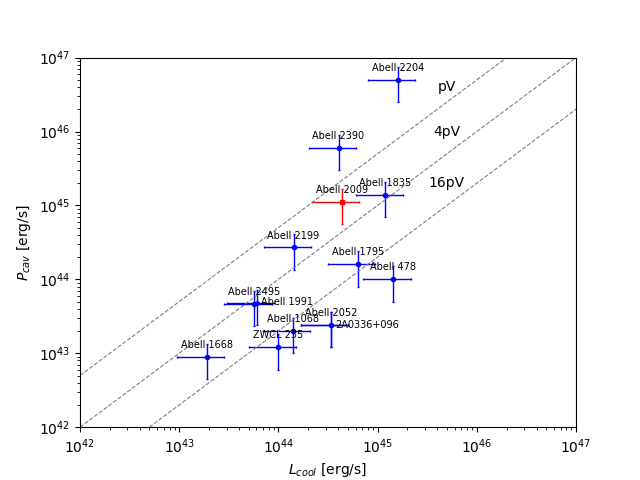}
        \label{}
    \end{subfigure}
    \begin{subfigure}[b]{0.44\textwidth}
        \centering
        \includegraphics[width=\textwidth]{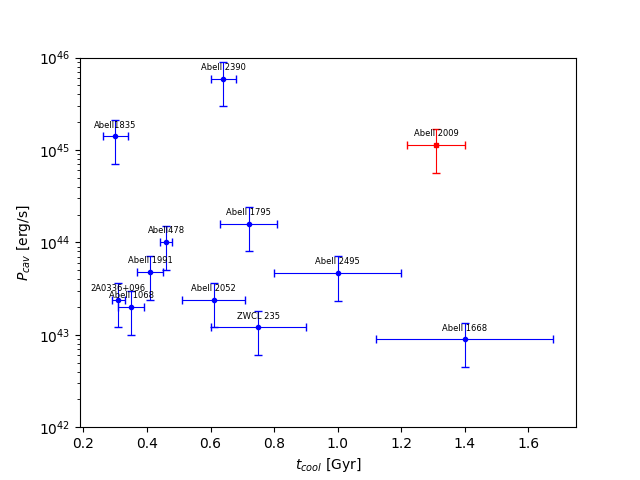} 
        \label{}
    \end{subfigure}
    
    \vspace{0.4cm}
    
    \begin{subfigure}[b]{0.44\textwidth}
        \centering
        \includegraphics[width=\textwidth]{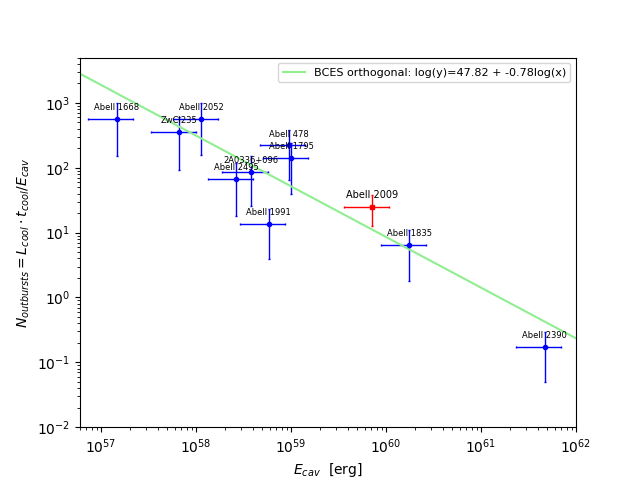} 
        \label{}
    \end{subfigure}
    \begin{subfigure}[b]{0.44\textwidth}
        \centering
        \includegraphics[width=\textwidth]{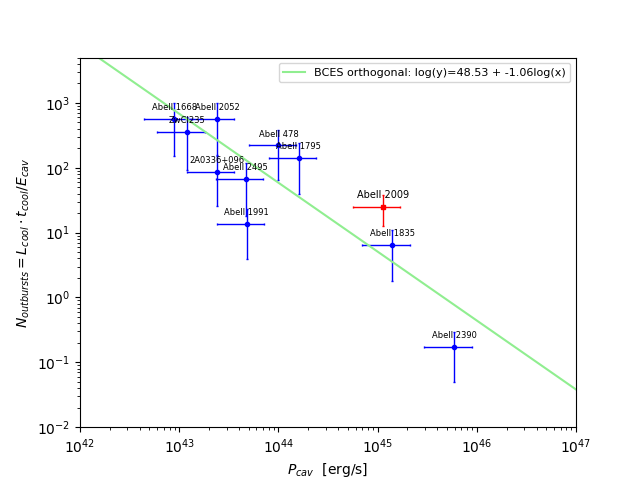} 
        \label{}
    \end{subfigure}
    
    \caption{Correlation between ICM and AGN properties for the galaxy clusters reported in Tab.\ref{samplevalues}. Upper left panel: Cavities power as function of the X-ray luminosity inside the cooling radius of each galaxy cluster. The diagonal lines indicate P$_{cav}$ = L$_{cool}$ assuming different values for the adiabatic index $\gamma$. All the uncertainties are set to 50\% for consistency. Upper right panel: Cavity power as a function of the cooling time at 12 kpc from the clusters center. All the uncertainties on the cavity power are set to 50\%, while the errors associated with the cooling time are reported in Tab.\ref{samplevalues}. Lower left panel: Number of outbursts required to counterbalance the radiative losses of the cooling gas as a function of the cavity energy. The green line represents the best-fit line derived using the BCES orthogonal regression method. All the uncertainties on the cavity energy and cooling luminosity are set to 50\%, while the errors associated with the cooling time are reported in Tab.\ref{samplevalues}. Lower right panel: Number of outbursts required to counterbalance the radiative losses of the cooling gas as a function of the cavity power. The green line represents the best-fit line derived using the BCES orthogonal regression method. All the uncertainties on the cavity energy, cavity power and cooling luminosity are set to 50\%, while the errors associated with the cooling time are reported in Tab.\ref{samplevalues}.}
    \label{sample}
\end{figure*}

The cavity power, P$_{cav}$, representing the mechanical energy injected into the hot gas by the AGN outburst, can then be directly compared to the gas luminosity within the cooling radius, L$_{cool}$, representing the radiative gas losses due to cooling.
As we show in Fig.\ref{sample} (upper left panel), cavity power scales in proportion to the cooling X-ray luminosity, albeit with considerable scatter.
This is a confirmation of the well-known relation between heating power and cooling losses found in several other works\footnote{We note that the definition of cooling luminosity of \citealt{2004ApJ...607..800B} and \citealt{2006ApJ...652..216R} differs from later studies, including ours. Specifically, these early works investigated the relation between $P_{cav}$ and $L_{X} - L_{spec,X}$, where $L_{X}$ and $L_{spec,X}$ are the luminosities inside the cooling region of the {\ttfamily apec} and {\ttfamily mkcflow} components, respectively, whereas subsequent studies have focused on $P_{cav}$ versus $L_{X}$ alone (e.g., \citealt{2012MNRAS.427.3468B}, \citealt{2013ApJ...777..163H}, \citealt{2023AandA...670A..23U}, \citealt{2023ApJ...948...49R}). The reason for using $L_{X}$ alone as the estimate of the cooling luminosity is mainly due to the limited spectral resolution of Chandra CCDs, which prevents accurate measurements of cooling rates in large spatial extraction regions, and to the fact that early estimates indicate that $L_{spec,X}$ is a small fraction ($\sim$10\% or less) of $L_{X}$ (e.g., \citealt{2008MNRAS.385.1186S}). In any case, we note that this potential offset given by neglecting $L_{spec,X}$ is  absorbed within the conservatively large systematic uncertainty ($\Delta L_X / L_X$ = 50\%) assumed in our analysis.} (e.g., \citealt{2004ApJ...607..800B}, \citealt{2006ApJ...652..216R}, \citealt{2012ARA&A..50..455F}).\\
We note that Abell~2204 has a significantly higher value for the cavities power with respect to the other galaxy clusters of the sample, lying well above the linear relation P$_{cav}$ = L$_{cool}$. This is mainly driven by the presence, according to \cite{2009MNRAS.393...71S}, of a giant cavity with a very large mechanical power in the cluster, which dominates the total energy budget. Similarly, Abell 2390 also lies above the same line, though with a smaller offset. This behavior indicates that the energy provided by its outburst exceeds the amount needed to counteract cooling.\\
Even though our selection of cool cores does not constitute a complete sample, it is likely representative of systems where AGN feedback is ongoing and as such, we expect a similar P$_{cav}$ - L$_{cool}$ scaling to hold in other similar systems. In this respect, we note that a linear scaling relation between these quantities was confirmed to exist even in complete samples of galaxy clusters (see the study of \citealt{2012MNRAS.427.3468B}).

\subsubsection{Cavity power versus cooling time}
We also examine the possible correlation between the power of the AGN's outburst and the central gas cooling time. 
From the point of view of the feeding side of the self-regulated loop, a short gas cooling time is able to supply the central AGN with a strong inflow of cool gas, possibly leading to more powerful mechanical outbursts, so an anticorrelation may be expected \citep{2006MNRAS.373..959D}. On the other hand, from the point of view of the feedback side of the self-regulated loop, long cooling times could be the result of the heating effect of a highly powerful outburst on the surrounding gas, so a direct correlation may be expected. However, as we show in Fig.\ref{sample} (upper right panel), no significant correlation is visible. For example, outbursts with power $\sim 10^{43} \text{erg s}^{-1}$ can be found in X-ray atmospheres whose central cooling time can vary by an order of magnitude, between about 0.2 and 2 Gyr. More insights on the P$_{cav}$ - t$_{cool}$ relation can be derived by considering that the two quantities reflect processes with different timescales; the cavities age ($\approx$ tens of Myr) is typically shorter than the time needed by the gas to cool ($\approx$ 1 Gyr). It is also evident that despite the highest energetic AGN outbursts, the gas cooling time remains below $\sim$ 1.5 Gyr in the central regions. In this respect, \cite{2012NJPh...14e5023M} pointed out that the most powerful AGN outbursts do not significantly raise the central gas temperature or lower the central gas density. On the other hand, they increase the central gas entropy through a gentle heating process.

\subsubsection{Outburst occurrence versus outburst energy and power}
We investigate the outburst occurrence by considering the quantity L$_{cool} \times$ t$_{cool}$ / E$_{cav}$, that represents the number of AGN outbursts of energy E$_{cav}$ required to counterbalance the radiative losses of the cooling gas (assuming a duty cycle = 1). As it can be seen in Fig.\ref{sample} (lower left and right panels), the L$_{cool} \times$ t$_{cool}$ / E$_{cav}$ quantity is plotted against the cavity energy, E$_{cav}$, and the cavity power, P$_{cav}$, respectively.  In both cases, the data are fitted using the BCES orthogonal regression method \citep{1996ApJ...470..706A}, revealing two linear anti-correlations. For the cavity energy plot, the slope of the best-fit line is $-0.78\pm 0.12$, with Pearson and Spearman correlation coefficients of $-0.92$ and $-0.79$, respectively. Similarly, for the cavity power plot, the slope is $-1.06\pm 0.26$, and the corresponding Pearson and Spearman coefficients are $-0.88$ and $-0.78$. In general, these plots indicate that more energetic outbursts and more powerful outbursts require a lower frequency in order to counterbalance the radiative losses due to cooling.\\
This is clearly exemplified by Abell 2390, that while lying well above the 4pV line in Fig.\ref{sample} (top left panel), also requires about 0.1 such outbursts per cooling time to counterbalance the radiative losses of its hot gas halo (Fig.\ref{sample}, lower panels). As explained in \citealt{2004ApJ...607..800B}, clusters that require less than 4pV will be reasonably supplied with energy in the cavities to balance cooling, while for clusters requiring more than 16pV per cavity, it is more difficult to balance cooling, if no other forms of heating are present. Indeed, as it can be observed in Fig.\ref{sample} (lower panels), clusters with a higher P$_{cav}$ value require a smaller amount of AGN outburst to balance the cooling.\\
A similar result is obtained in the study of \cite{2014MNRAS.442.3192V}, where the cavity heating time was defined as the amount of time that the cavity energy is able to offset radiative losses, which is E$_{cav}$ / L$_{cool}$. Comparing the cavity heating time with the outburst interval of 13 different systems, the authors derived that more energetic outbursts offset cooling with longer outburst intervals. Moreover, the outburst interval is often shorter than the correspondent cavity heating time, suggesting that heating continually compensates for radiative cooling through the cluster's lifetime.\\
\\
To conclude, we have investigated the relationship between AGN heating and ICM cooling properties in a collection of 18 clusters, adding our results on Abell~2009 to the literature work on the remaining clusters, performed using different methods. However, we note that a homogeneous analysis of larger samples would be necessary to better investigate and confirm the relationships discussed in this work.

\section{Conclusions}
In this work, we presented {\it Chandra} and 1.5 GHz VLA observations of the galaxy cluster Abell~2009 by performing in particular a detailed X-ray analysis of its ICM for the first time. \\
The main results we obtained are briefly summarized below:

   \begin{itemize}
      \item The BCG has a total 1.5 GHz flux density of 23 $\pm$ 1 mJy, with 5.8 $\pm$ 0.3 mJy coming from the central bright core. Its extended radio lobes present a symmetric butterfly-shaped morphology. The radio galaxy in the N direction has a total flux density of 41 $\pm$ 1.6 mJy, including the core emission, and is a cluster member of Abell 2009. Located approximately 130 kpc from the center of Abell 2009, this radio source may be interacting with the cluster's ICM.
      \item The global properties of the ICM of Abell~2009 are kT = 6.41 $\pm$ 0.17 keV and Z = 0.59 $\pm$ 0.07 Z$_{\odot}$. The morphological analysis confirmed the cool-core nature of Abell 2009, since its surface brightness profile is peaked at center and is best-fitted with a double $\beta$ model. From the spectral analysis, the cooling radius results to be $r_{cool}=87.5\pm 15.7$ kpc, within which the cooling time is lower than 7.7 Gyr. The X-ray luminosity inside the cooling region is $L_{cool}$ = $(4.4\pm 0.1)\times 10^{44}$ erg s$^{-1}$.
      \item As found for several other cool-core clusters, we confirm the mismatch in Abell~2009 between the theoretical, global mass inflow rate ($\dot{M}_{glob}= (316 \pm 16)$ M$_{\odot}$/yr) expected in the standard cooling flow model and the spectral mass inflow rate ($\dot{M}_{spec} \leq 19.4$ M$_{\odot}$/yr). This discrepancy is consistent with the presence of a heating mechanism that can reduce the efficiency of the cooling process in the central region of the cluster.
      \item No clear depressions are detected in the X-ray image, possibly due to the relatively short exposure. Guided by the radio images under the assumption that any underlying X-ray cavity is traced by radio bubbles, we measure the properties of possible cavities corresponding to the radio lobes of the BCG. The N lobe and S lobe have ages of $\approx$ 27 Myr and $\approx$ 14 Myr, respectively. The lobes total power is $\approx 10^{45}$ erg s$^{-1}$. Comparing this value with the X-ray luminosity inside the cooling region, it can be inferred that the mechanical power associated to the activity of the central AGN is able to counterbalance the cooling process of the ICM.
      \item Finally, we briefly investigate the general properties of the parent sample of galaxy clusters with high X-ray flux and high H$\alpha$ luminosity from which Abell 2009 was selected. We discuss the lack of correlation between cavity power and cooling time as a result of the feedback from cluster-central AGN being gentle and not disruptive. Moreover, we find indications that more energetic and powerful outbursts require a lower frequency, compared to less energetic ones, to balance the radiative losses resulting from gas cooling.
   \end{itemize}

\begin{acknowledgements}
      This paper and related research have been conducted during and with the support of the Italian national inter-university PhD programme in Space Science and Technology - University of Trento / IUSS Pavia.\\
      FU and MG acknowledge the financial contribution from contract PRIN 2022 - CUP J53D23001610006.\\
      The authors thank the anonymous reviewer for providing useful feedback on this manuscript.
\end{acknowledgements}

%
%
\balance
\bibliographystyle{aa}
\bibliography{aanda}

\appendix
\onecolumn
\section{Additional table}
In this Section, we report the table with the best-fit results of the projected spectral analysis (see Subsec.\ref{proj}). 
\begin{table*}[h]
\centering
\caption{Best-fit results of the projected radial analysis of Abell 2009.}
\begin{tabular}{ccccccc}
\hline
  R$_{in}$ - R$_{out}$ [kpc]&kT [keV] & Z [Z$_{\odot}$]& norm [10$^{-3}$]& Net counts &$\chi^2$/d.o.f.  \\
 \hline
  3.9 - 37.4& 4.78$^{+0.28}_{-0.25}$ & 1.07$^{+0.23}_{-0.21}$ & 1.52$^{+0.07}_{-0.07}$ & 2892 (99.4\%)& 66.98/85 (0.79)\\
\hline
37.4 - 67.6& 6.55$^{+0.59}_{-0.46}$ & 0.64$^{+0.25}_{-0.23}$ & 1.48$^{+0.06}_{-0.06}$ & 2612 (98.6\%)& 84.08/76 (1.11)\\
\hline
67.6 - 98.9& 6.95$^{+0.64}_{-0.54}$ & 0.59$^{+0.22}_{-0.21}$ & 1.41$^{+0.05}_{-0.05}$ & 2473 (97.5\%) & 79.16/74 (1.07)\\
\hline
98.9 - 135.8& 7.08$^{+0.67}_{-0.55}$ & 0.96$^{+0.29}_{-0.27}$ & 1.38$^{+0.06}_{-0.06}$ & 2556 (96.2\%) & 65.20/78 (0.84)\\
\hline
135.8 - 182.7& 7.12$^{+0.84}_{-0.59}$ & 0.28$^{+0.18}_{-0.18}$ & 1.63$^{+0.05}_{-0.05}$ & 2745 (93.9\%) & 81.19/82 (0.99)\\
\hline
182.7 - 244.7& 7.17$^{+0.62}_{-0.56}$ & 0.78$^{+0.26}_{-0.24}$ & 1.51$^{+0.06}_{-0.06}$ & 2697 (89.7\%) & 83.98/82(1.02)\\
\hline
244.7 - 326.9& 7.34$^{+0.91}_{-0.72}$ & 0.38$^{+0.24}_{-0.24}$ & 1.46$^{+0.06}_{-0.06}$ & 2412 (81.3\%) & 74.51/79 (0.96)\\
\hline
326.9 - 434.2& 6.85$^{+0.82}_{-0.66}$ & 0.37$^{+0.28}_{-0.26}$ & 1.23$^{+0.06}_{-0.06}$ & 2128 (68.8\%) & 75.99/77 (0.99)\\
\hline
\end{tabular}
\tablefoot{Best-fit results of the elliptical projected radial spectrum of Abell 2009. In the first column the internal and outer radii of each elliptical annulus are reported (these radii correspond to the semi-major axes of the annuli). The second, third, and forth columns are for the temperature, metallicity, and normalization. The last two columns report the net photon counts and their percentage with respect to the total counts, and the value for the reduced $\chi^2$ of each fit.}
\label{projected}
\end{table*}

\end{document}